# Hierarchical Multi-Objective Optimization for Precise Performance Design of Closed-Chain Legged Mechanisms

Long Guo[1, 2], Ying Zhang[1]*, Qi Qin[1], Guanjun Liu[3], Hanyu Chen[2], Yan-an Yao[1]

[1]Institute of Intelligent Robots and Systems, School of Mechanical, Electronic and Control Engineering, Beijing Jiaotong University, Beijing, 100044, P.R. China
[2]Department of Mechanical Engineering, Tsinghua University, Beijing, 100084, P.R. China
[3]School of Mechanical and Electrical Engineering, Shanxi Datong University, Shanxi, 037003, P.R. China
*Corresponding Author (email: yzhang2@bjtu.edu.cn)

*Abstract*: Over the past decades, the performance design of closed-chain legged mechanisms (CLMs) has not been adequately addressed. Most existing design methodologies have predominantly relied on trajectory synthesis, which inadvertently prioritizes less critical performance aspects. This study proposes a hierarchical multi-objective optimization strategy to address this limitation. First, the numerical performance-trajectory mapping is derived based on a foot-ground contact model, aiming to decouple the performance characteristics. Subsequently, a hierarchical optimization strategy is employed for two CLM design scenarios: In trajectory shape-constrained scenarios, a coarse-to-fine optimization process, integrating Fourier descriptors, refines the design from overall shape to local features. In scenarios without trajectory shape constraints, a stepwise optimization process is proposed for reconfigurable CLMs to transition from primary motion to auxiliary motion. The robustness of the proposed design strategy is validated across three configurations and seven algorithms. The effectiveness of the proposed design strategy is verified by comparison with other existing CLM design methods. The applicability of the proposed strategy is confirmed through simulation and prototype experiments. The results demonstrate that the hierarchical strategy effectively addresses the challenges of precise performance design in CLMs. Our work provides a general framework for the CLM design and offers insights for the optimization design of other closed-chain linkages.

## 1  Introduction

Closed-chain legged mechanisms (CLMs) [1-3] are robotic leg structures where the links form one or more closed kinematic loops. Low-degree-of-freedom (DoF) CLMs are often regarded as less adaptable and flexible than open-chain legs due to their limited trajectory domain. However, recent advancements in dimensional synthesis and reconfigurable design, in conjunction with control strategies based on environmental perception and reinforcement learning [4], have emerged as promising avenues for enhancing the load-bearing capacity of low-DoF CLMs while simultaneously improving their terrain adaptability. This study aims to solve the precise performance design problem of CLM through hierarchical multi-objective optimization strategies.

In CLM design scenarios with trajectory shape constraints, the transformation of CLM designs into trajectory synthesis problems has emerged as a prevailing approach [5]. The bench trajectory (BT) directly corresponds to the design target for this particular problem, signifying the closed-loop foot-end trajectory when the trunk is fixed. Wu [6] proposed a methodology that utilizes six points to delineate the desired trajectory, formulating a discrete points deviation function. This function was subsequently refined using the gradient-based optimization algorithms to generate an elliptical foot trajectory. Wang [7] employed a heuristic algorithm to optimize a trajectory with 13 points, successfully approximating a cycloid trajectory on a two-DoFs CLM. Several studies have worked on improving the accuracy of heuristic algorithms for trajectory synthesis. In a seminal study, Wu [8] proposed an overall shape-based function via Fourier series, which was then combined with a genetic algorithm to reproduce a symmetric crescent trajectory of an 8-bar CLM. Building on these earlier contributions, Guo [9] proposed a novel combined method that enhance the accuracy of trajectory synthesis. This method integrates a multi-objective heuristic algorithm with a gradient-based algorithm, facilitating the generation of a boot-shaped foot trajectory in 4-bar to 14-bar linkages. Despite these advancements, trajectory synthesis still faces challenges in executing certain tasks, such as generating cycloidal trajectories [10, 11] that exhibit zero center-of-mass fluctuation and ensure soft landing. Cycloidal trajectories are prevalent in multi-DoFs serial legged robots, and attempts to produce these trajectories using single-DoF CLMs have yielded unsatisfactory results in terms of shape and trajectory performance [12]. This shortfall can be attributed to the majority of trajectory synthesis methods, which primarily emphasize the overall shape similarity. Examples include the Fourier series method [13-15], the wavelet series method [16, 17], and the B-spline curve method [18, 19]. These methods neglect to consider the constraints or optimization of detailed trajectory features. Consequently, the integration of a comprehensive trajectory synthesis approach with the optimization of local trajectory features, through a hierarchical optimization strategy, has the potential to achieve precise performance design for CLMs.

In CLM design scenarios without trajectory shape constraints are frequently observed in reconfigurable legs [20, 21]. These mechanisms are characterized by the possession of additional DoFs, thereby enabling alterations in their geometric or topological relationships. This adaptability facilitates the switching of working modes in response to varying environmental demands. Such designs prioritize the parameters of WT, which refers to the open foot trajectory under foot-ground constraints. Ruan [22] contributed to this field by incorporating a rotating motor into the trunk of a CLM. This modification increased the initial obstacle clearance height of walking trajectory (WT) from approximately 100 mm to 250 mm when the trunk pitch angle was adjusted from 0° to 25°. Wei [23] introduced an approach of adjusting the frame length, thereby achieving a continuous modification of the BT. This adjustment escalated the crossing height from 85 mm to 290 mm. Wu [24] focused on the first loop of a Watt leg, where altering the rod length enhanced the average obstacle-crossing height from 153 mm to 283 mm. Zhao [25] developed a biomimetic quadruped robot capable of executing various locomotion modes, such as rotating and stair-climbing gaits, through a single-loop metamorphic mechanism. Wu [26] coupled the CLM with an adjustable trunk that consists of a double DoFs single-loop mechanism, forming a CLM that can change obstacle-crossing capability by trunk deformation. The complexity of dimensional coupling and constraint relationships poses significant challenges. Moreover, current design methodologies for reconfigurable CLMs predominantly utilize sensitivity analysis methods [17]. These methods adjust only one or a few parameters that most influence performance, limiting the precise performance design of reconfigurable CLMs. In summary, reconfigurable CLMs demonstrate considerable performance potential; however, to the best of our knowledge, there is currently no dimensional design method specifically tailored for reconfigurable CLMs, let alone one that enables precise performance design.

From a methodological perspective, CLM performance design methods involve optimization [9], analysis [27, 28], and data-driven methods [29, 30]. Optimization methods have been widely applied in engineering design [31-34], which can address complex constraints in CLM designs [35, 36], such as linkage defects (e.g., crank, circuit, and branch issues [37-41]), preventing backward walking, and ensuring proper landing and take-off angles to avoid slipping. However, existing optimization methods in CLM design frequently encounter difficulties in converging to optimal solutions when only approximate feasible solutions exist. This challenge stems from four primary issues: the absence of quantifiable metrics to assess walking performance, the challenge in decoupling performance metrics, the inability to accurately estimate performance limits, and the complexity in prioritizing performance objectives.

Based on the above analysis, this study initially derives the numerical mapping between trajectory performance and CLM motion performance using a foot-ground contact model, aimed at reducing the coupling between performance metrics. Subsequently, a hierarchical multi-objective optimization approach is applied to address two challenging design tasks: Section 3 designs a CLM capable of replicating a compound cycloidal trajectory with minimal fluctuations during the stance phase and soft landing. Section 4 targets the design of a reconfigurable CLM with precisely controlled obstacle-crossing performance across various modes.

## *2 Numerical Performance-Trajectory Mapping*

This section establishes a numerical mapping between the trajectory and CLM motion performance based on the foot-ground contact model, reducing the strong coupling inertia of the performance parameters and setting the foundation for the subsequent hierarchical strategy.

2.1 Features points of BT

As shown in Fig. 1, the CLM serves as the basic unit. Two CLMs form a biped module [42] with a symmetric gait. Two biped modules create a quadruped robot [8] or module [43] with a trot gait. Combining multiple quadruped modules forms a multi-leg robot with a stable supporting polygon, ensuring stability based on Zero Moment Point theory.

Fig. 2 illustrates eight feature points used to characterize BT. The Assur Group Method is employed for calculating discrete points of BTs. The trajectory period is denoted as *T*. BT is sampled uniformly for discretization. The BT vector is represented as $C_b(t) = [x_b(t), y_b(t)]'$, where $x_b$ and $y_b$ denote the respective coordinates. The feature points of BT includs the landing point $C_b(t_1)$, take-off point $C_b(t_2)$, lowest point $C_b(t_3)$, left-most $C_b(t_4)$, right-most points $C_b(t_5)$, local lowest point $C_b(t_6)$, and two highest point $C_b(t_7)$ and $C_b(t_8)$. For both bipedal symmetric and quadrupedal trot gaits, $C_b(t_1)$ and $C_b(t_2)$ must meet three criteria: a 180-degree difference in crank angles, implying $t_4 = t_2 + T/2$; equal y-coordinates: $y_b(t_1) = y_b(t_2)$; and $x_b(t_1)$ being less than $x_b(t_2)$: $x_b(t_1) < x_b(t_2)$. The positions of $t_1$ and $t_2$ are determined based on these conditions and the Assur Group Method. The remaining feature points are identified by analyzing changes in the velocity vector direction and by ordering and indexing the $C_b(t)$.

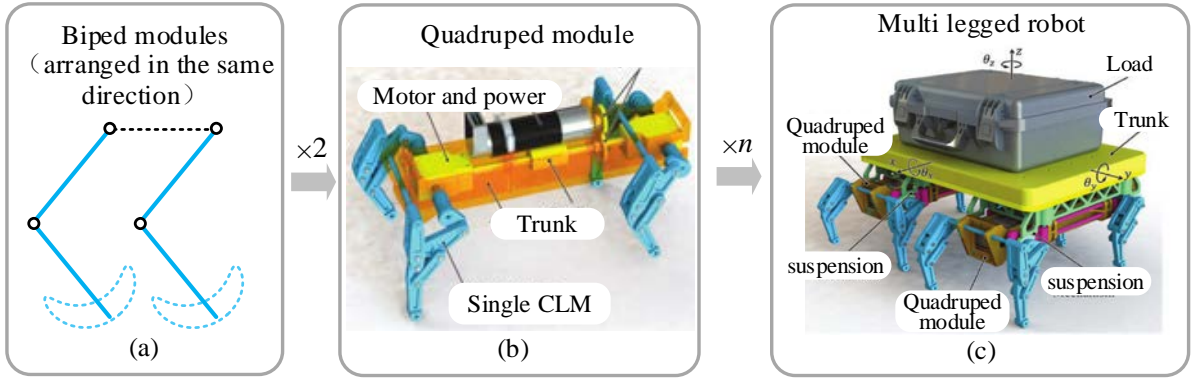

Fig.1 Process of constructing multilegged robot with bipedal module: (a) Biped module, (b) quadruped module [43], and (c) multi legged robot [43].

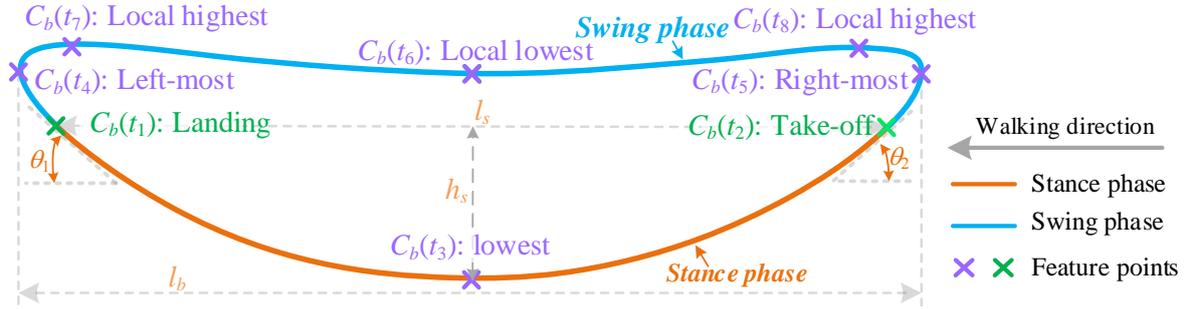

Fig. 2 Feature points of the bench trajectory.

### 2.2 BT to WT conversion

The WT is derived from the BT based on the foot-ground contact model (assuming no relative sliding). Let $C_m(t)$ be defined as the discretized coordinate vector of WT, represented as $[x_m(t), y_m(t)]'$. Here, $x_m$ and $y_m$ denote the trajectory coordinates. Let $C_b^K$ and $C_m^K$ represent the BT and WT of Leg $K$, respectively. As shown in Fig. 3, the two legs have a phase difference of $T/2$ under the symmetric gait [44]. Taking the take-off point of Leg $A$ as the time origin, after $\Delta t$ seconds ($0 < \Delta t \leq T/2$), the $C_b^A$ and $C_b^B$ relative to $C_b^A(t_2)$ and $C_b^B(t_1)$ are recorded as $\Delta C_b^A$ and $\Delta C_b^B$. $C_m^A$ is the composition of the motions of $C_b^A$ and $C_b^B$:

$$C_m^A(t_2 + \Delta t) = C_b^A(t_2) - \Delta C_b^B(\Delta t) + \Delta C_b^A(\Delta t) \qquad 0 < \Delta t \leq \frac{T}{2} \tag{1}$$

where,

$$\Delta C_b^B(t) = C_b(t - t_2 + t_1) - C_b(t_1) \tag{2}$$

$$\Delta C_b^A(t) = C_b(t) - C_b(t_2) \tag{3}$$

Eqs. (1-3) are combined and transformed into a normalized expression of $C_m$ within one cycle:

$$C_m(t) = \begin{cases} C_b(t) + C_b(t_1) - C_b(t + t_1) & 0 \leq t < T/2 \\ 0 & T/2 \leq t < T \end{cases} \tag{4}$$

From Eq. 4, we know that:

$$C_m(t_1) = 2C_b(t_1) \tag{5}$$
$$C_m(t_2) = C_b(t_2) \tag{6}$$

This implies that the BT and WT coincide at the take-off point, and the WT step length equals twice the stance phase length of the BT.

### 2.3 Relative positions of multiple WTs

Fig. 4 shows the Z-direction view of two WTs, located at the take-off and landing points. In the quadrupedal trot gait, diagonal pairs of legs synchronize their movements. The four legs are designated as Legs $A$ to $D$, and individual local coordinate systems, labeled as $\{A\}$ to $\{D\}$, are established at the rotation centers of these four legs. Let $^KC$ denote the expression of the trajectory within $\{K\}$. The difference in the $x$-coordinates of the foot of Legs $M$ and $N$ is expressed as $\Delta x_{M,N}$. The difference in the $x$-coordinates of legs $A$ and $B$, $\Delta x_{A,B}$, can be expressed as:

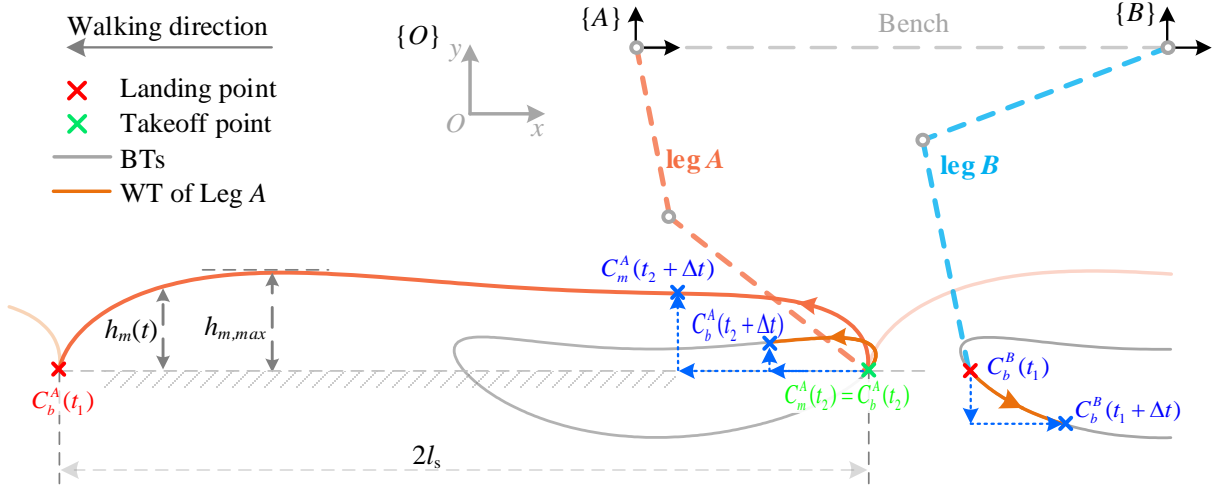

Fig. 3 BT to WT conversion.

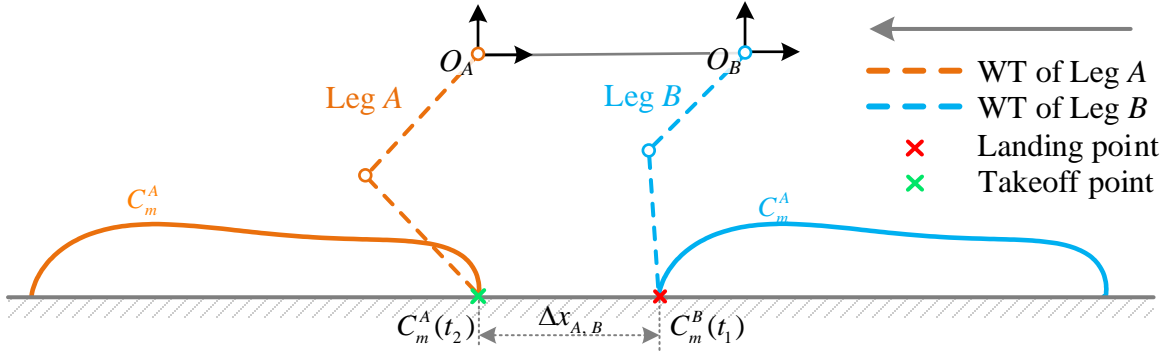

Fig. 4 Z-direction view of the two WTs.

$$\begin{bmatrix} \Delta x_{A,B} \\ 0 \end{bmatrix} = {}^A C_m^B(t_1) - {}^A C_m^A(t_2) = {}^B C_m^B(t_1) + O_B - O_A - {}^A C_m^A(t_2) \tag{7}$$

Similarly, it can be concluded that:

$$\begin{bmatrix} \Delta x_{B,C} \\ 0 \end{bmatrix} = {}^B C_m^C(t_2) - {}^B C_m^B(t_1) = {}^C C_m^C(t_2) - {}^B C_m^B(t_1) \tag{8}$$

$$\begin{bmatrix} \Delta x_{C,D} \\ 0 \end{bmatrix} = {}^C C_m^D(t_1) - {}^C C_m^C(t_2) = {}^D C_m^D(t_1) + O_4 - O_3 - {}^C C_m^C(t_2) \tag{9}$$

where, according to Eq. (5-6), it can be inferred that:

$$\begin{cases} {}^A C_m^A(t_2) = {}^C C_m^C(t_2) = C_b(t_2) \\ {}^B C_m^B(t_1) = {}^D C_m^D(t_1) = 2C_b(t_1) \end{cases} \tag{10}$$

By solving Eqs. (7-10), the values of $\Delta x_{A,B}$, $\Delta x_{B,C}$, and $\Delta x_{C,D}$ can be obtained, which allow us to determine the relative position relationship of the WTs in both the biped (see Fig. 5 (a)) and quadruped modules (see Fig. 5 (b)).

2.4 Performance-trajectory mapping

Based on the BT and WT, three types of performance parameters are quantified: body fluctuate, ground impact, and obstacle-crossing ability.

2.4.1 Body fluctuate

Let $h_s$ denote the maximum value of the stance trajectory fluctuations. The leg length variation influences the height of the body frame as the supporting foot makes contact with the ground. Therefore, the body fluctuation in the swing phase is equal to the BT fluctuation in the stance phase.

$$h_s = y_b(t_1) - \min(y_b) \tag{11}$$

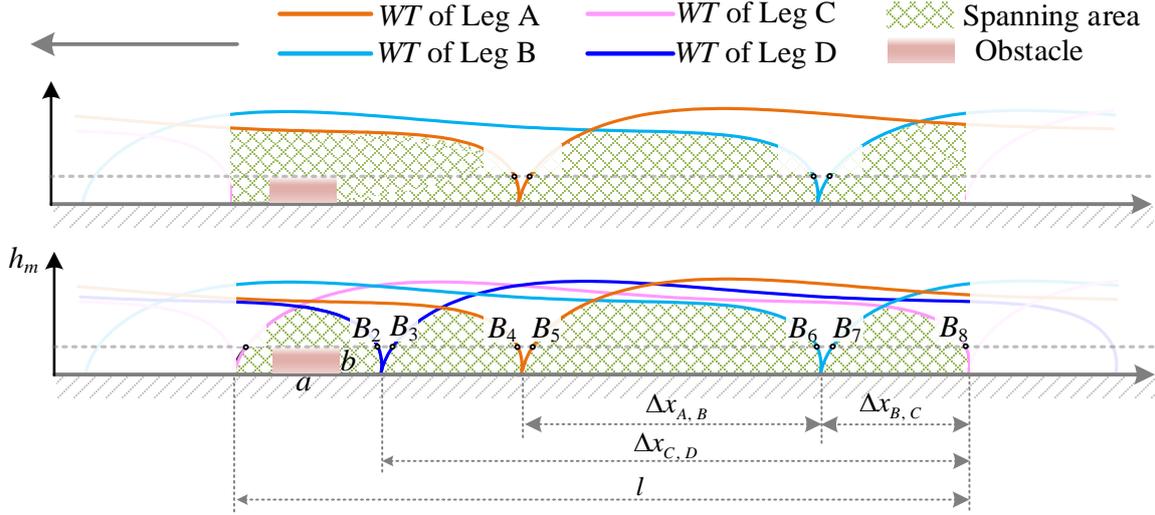

Fig. 5 Distribution of walking trajectories, (a) bipedal effective crossing region and (b) quadrupedal effective crossing region.

The straightness $S$ of the stance phase is a dimensionless parameter that indicates the smoothness of the body frame during motion. It is defined as the ratio of the longitudinal fluctuation of the body to the length of the stance phase, which eliminates the effect of the mechanism size. The formula is given by Eq. 12.

$$S = h_s / l_s \times 100\% \quad (12)$$

where, $l_s$ represents the length of the stance phase of BT.

2.4.2 Ground impact

The impact force model of ground impact can be equivalent to a spring-damper system [45], and the magnitude of the impact force $F_n$ can be expressed as:

$$F_n(t) = \lambda |y|^p \dot{y} + k |y|^p \quad (13)$$

where, $\lambda$ denotes hysteresis damping factor and $k$ denotes the spring constant. $p$ is the power exponent, and $y$ and $\dot{y}$ are the penetration depth and velocity, respectively. When there is no relative sliding between foot and ground, $\dot{y}$ is positively correlated with the foot velocity. Hence, the magnitude of the vertical velocity at the trajectory landing point characterizes the impact $I$ of the foot on the ground.

$$I = \dot{y}_b(t_1) \sin \theta_1 \quad (14)$$

where, $\theta_1$ denotes the grounding angle, which is the angle between the tangent of the BT landing point and the positive $x$-axis. $\theta_1$ can be estimated by the median difference:

$$\theta_i = \arctan\left(\left|\frac{y_b(t_i + \Delta t_{min}) - y_b(t_i - \Delta t_{min})}{x_b(t_i + \Delta t_{min}) - x_b(t_i - \Delta t_{min})}\right|\right) \quad (15)$$

2.4.3 Obstacle-crossing ability

$y_m$ indicates the height of obstacles that a CLM can cross. Let $h_m$ be the maximum crossing height:

$$h_m = \max(y_m) \quad (16)$$

Let $\bar{h}$ be the mean definite integral of $y_m$ in swing phase, which represents the average height that a CLM can cross.

$$\bar{h} = \frac{2}{T} \int_{t_2}^{t_2 + \frac{T}{2}} y_m dt - \min(y_m) \quad (17)$$

Similarly, we can also calculate the probability $\psi(a, b)$ of CLMs crossing an obstacle with length $a$ and height $b$, as shown in Eq. 18. Fig. 5 shows the effective crossing region in green. The effective crossing region is where CLM can safely and efficiently cross the obstacle without colliding or losing balance. The shape and size of the effective crossing region depend on the obstacle geometry and the CLMs' gait. For brick-shaped obstacles, the CLMs can cross the obstacle if it is within the bipedal effective crossing region, as shown in Fig. 5 (a). For infinitely long obstacles, such as stairs, the obstacles need to be within the quadrupedal effective crossing region, as shown in Fig. 5 (b).

$$\psi_2(a,b) = \frac{l - B_4B_5 - B_6B_7 - 4a}{l}$$
$$\psi_4(a,b) = \frac{B_1B_2 + B_3B_4 + B_5B_6 + B_7B_8 - 4a}{l} \tag{18}$$

where, $\psi_2(a, b)$ and $\psi_4(a, b)$ denote the probabilities of crossing brick-shaped and step-shaped obstacles, respectively. $B_1 \sim B_8$ are defined as the points where WTs intersects the horizontal line $h_m = b$, as illustrated in Fig. 5.

In this section, we have quantified three key performance parameters for general legged robots based on BT and WT, namely the body fluctuations, the ground impact, and the obstacle-crossing ability. These three are the most common for CLMs and will be used later. Some studies also have emphasized the symmetry [8] or other trajectory features, which we haven't listed them all here.

## 3 CLM design with trajectory shape constraints

This section presents the framework of CLMs design with a desired composite cycloidal trajectory using a hierarchical optimization strategy. An initial solution is generated using a Fourier series-based method, followed by trajectory refinement through optimization from the overall shape to local details. Simulations and a prototype confirm the design's feasibility. The validation on a variety of linkage configurations and seven latest algorithms demonstrates the strategy's effectiveness. The comparison with single-objective optimization and other existing CLM design methods demonstrates the strategy's superior capabilities.

3.1 Predesign

Research has shown the efficacy of trajectory planning in open-chain legs [46-48], leading to various foot trajectory proposals, such as the compound cycle [49], elliptic [50], polynomial [51], and spline curve [52]. Notably, the composite cycloidal trajectory, known for mitigating body fluctuation and touchdown impact, has yet to be successfully implemented in CLMs. The analytical expression of the compound cycloid is as follows:

$$\begin{cases} x(t) = L(\frac{t}{T} - \frac{1}{2\pi}\sin\frac{2\pi t}{T}) \\ y(t) = -H(\frac{1}{2} - \frac{1}{2}\cos\frac{2\pi t}{T}) \end{cases} \tag{19}$$

where, $L$ denotes the step length, $H$ denotes the step height. In this study, we have chosen specific values for the parameters: $L = 300$ mm, $H = 100$ mm, and $T = 2$ s. The desired trajectory is shown in Fig. 6.

Trajectory synthesis based on our previous study [9] was performed firstly to provide an approximate solution. This approach involved establishing a multi-objective optimization function, which focuses on the discrepancy in normalized harmonic parameters (Fourier descriptor) between the planned BT and optimized BTs. In this study, the non-dominated Sorting Genetic Algorithm-II (NSGA-II) and Stephenson-I linkage (see Appendix A) were initially employed as the example. Then, six recently published multi-objective optimization algorithms in 2024 (CMOEMT [56], CMOES [57], DRLOS-EMCMO [58], IMTCMO [59], MCCMO [60], and MOEA/D-CMT [61]) and two alternative configurations (Watt-I and Stephenson-III) were tested for comparison. The population size was set as 200, and both the crossover fraction and mutation fraction were set to 1/8. The solution process is consistent with that in [9]. The trajectory synthesis results, labeled as $X_0$, are presented in Fig. 6 (a) and Appendix A. Key performance metrics included a body fluctuation ($h_s$) of 12.5 mm, a straightness ($S$) of 4.17%, and a ground impact velocity ($I$) of 49.8 mm/s at a driving frequency of 2 Hz. The trajectory's mean squared error (MSE) was calculated at 9.91 mm, where MSE = $\Sigma((x_{i,o} - x_i)^2 + (y_{i,o} - y_i)^2)^{1/2}$, $(x_{i,o}, y_{i,o})$ represents the coordinates of the optimized BTs.

3.2 Performance design based on hierarchical strategy

This task involves multiple conflicting parameters. To address this complexity, a hierarchical strategy was developed, which divides the following design problem into three subtasks:
*Subtask 1*: Overall shape and position optimization;
*Subtask 2*: Local features optimization;
*Subtask 3*: Fine-tuning the performance features.

3.2.1 *Subtask 1*: Overall shape and position optimization
This step focus on optimizing shape and position parameters to enhance overall similarity. Thus, the objective functions are outlined in Eq. 20.

$$\min. \begin{cases} f_1 = \text{MSE} + g(x) \\ f_2 = \begin{cases} (l-300)^2 + g(x) & 290 < l < 310 \\ 5 \times (l-290)^2 + g(x) & 290 > l \\ 5 \times (l-310)^2 + g(x) & l > 310 \end{cases} \\ f_3 = (y_h(t_3) + 100)^2 + g(x) \end{cases} \quad (20)$$

where, a penalty function $g(x)$ is used, which generates a very poor individual ($10^{10}$) if there are no kinematic solutions, effectively removing them from the optimization process over time. The function $f_1$ ensures the shape closely aligns the desired compound cycloidal trajectories, and $f_2$ and $f_3$ are used to define the trajectory's position.

The search range is set at a 20% scaling and ±20 mm adjustment of $X_0$. The NSGA-II algorithm was first utilized with a population size of 200 for this multi-objective optimization problem, with both the crossover fraction and mutation fraction set to 1/8. A sensitivity analysis using the One-at-a-Time method was performed to ensure the rationality of the selected parameters. The results, depicted in Fig. 7 (a), show the distribution of individuals at the end of the optimization process. An individual labeled $X_1$ (see Appendix B), located on the Pareto optimal frontier, is identified with an MSE of 5.09 mm (a 48.64% decrease), and values of $f_2$ and $f_3$ below 0.01 mm$^2$, as illustrated in Fig. 6 (b). This individual's trajectory is similar to the compound cycloidal trajectory in both shape and position. Notably, the stance phase fluctuation $h_3$ is reduced to 11.07 mm (a 11.44% decrease), and the straightness $S$ is 3.69%. The ground impact $I$ increases to 137.9 mm/s (a 2.77-fold increase) due to the absence of constraints.

3.2.2 *Subtask 2*: Local features optimization

Using the optimized shape and position parameters from *Subtask 1* as constraints, stance phase fluctuation $h_3$ and ground impact $I$ were further optimized. Adhering to the trajectory shape similarity constraint (Eq. 21), we redefine objective functions $f_1$ - $f_3$ from *Subtask 1* as constraints (Eq. 22), with values set at [6, 5, 5] based on prior findings. A 10% scaling and ±10 mm adjustment of $X_1$ is applied as the search range. The individuals with smaller values of $f_4$ are prioritized.

$$\min. \begin{bmatrix} f_4 \\ f_5 \end{bmatrix} = \begin{bmatrix} h_s + g(x) \\ I + g(x) \end{bmatrix} \quad (21)$$

$$\text{s.t.} \begin{cases} f_1 < 6 \\ f_2 < 5 \\ f_3 < 5 \end{cases} \quad (22)$$

The distribution of individuals in the final generation is visualized in Fig. 7 (b), with a noteworthy individual on the Pareto optimal frontier labeled as $X_2$ (see Appendix B), as shown in Fig. 6 (c). The performance improvement is significant: MSE is reduced to 4.8 mm (a 5.7% decrease), $h_3$ is reduced to 6.26 mm (a 43.45% decrease), and $I$ is reduced to 16.25 (a remarkable decrease of 88.22%).

3.2.3 *Subtask 3*: Fine-tuning the performance features

Based on the results of *Subtasks 1* and *2*, we fine-tune the trajectory performance within a small neighborhood. All previously optimized parameters (shape, position, and initial performance) are used as constraints to allow precise adjustments, aiming for the desired compound cycloidal trajectory. In this step, the objective functions from Eq. 21 are applied. A 5% scaling and ±5 mm adjustment of $X_2$ is applied as the search range. Additional constraints introduced as shown in Eq. 23. The individuals with smaller values of $f_5$ are prioritized.

$$\text{s.t.} \begin{cases} f_4 < 6.5 \\ f_5 < 20 \end{cases} \quad (23)$$

The distribution of individuals in the final generation is illustrated in Fig. 7 (c), with a specific individual on the Pareto optimal frontier denoted as $X_3$ (see Appendix B), as seen in Fig. 6 (d). The comparison of performance parameters of different individual is shown in Appendix B-1. After three optimization subtasks, The MSE is reduced to 3.97 mm (a 17.19% decrease), $h_3$ is reduced to 5.80 mm (a 7.35% decrease), and $I$ is reduced to 0.9 mm/s (a remarkable 94.46% decrease).

In summary, the CLM design of with compound cycloidal trajectory focused on achieving precise performance while adhering to shape constraints. This hierarchical strategy has successfully implemented the compound cycloidal trajectory within CLMs, achieving a 59.9% reduction in MSE, a 77.9% reduction in $h_3$, and a 98.2% reduction in $I$.

3.2.4 Decision-making criteria

The decision-making process for $X_i$ is comprised of two steps. First, Knee-Point Identification is employed to objectively select the most representative Pareto-optimal solutions, typically reducing the candidate pool to fewer than ten. Then, a small-scale manual selection process is employed, considering geometric symmetry, trajectory reproduction accuracy in the landing phase, and previously computed performance parameters. This decision-making criteria integrates objective filtering with expert judgment, ensuring robust solution quality while maintaining computational efficiency.

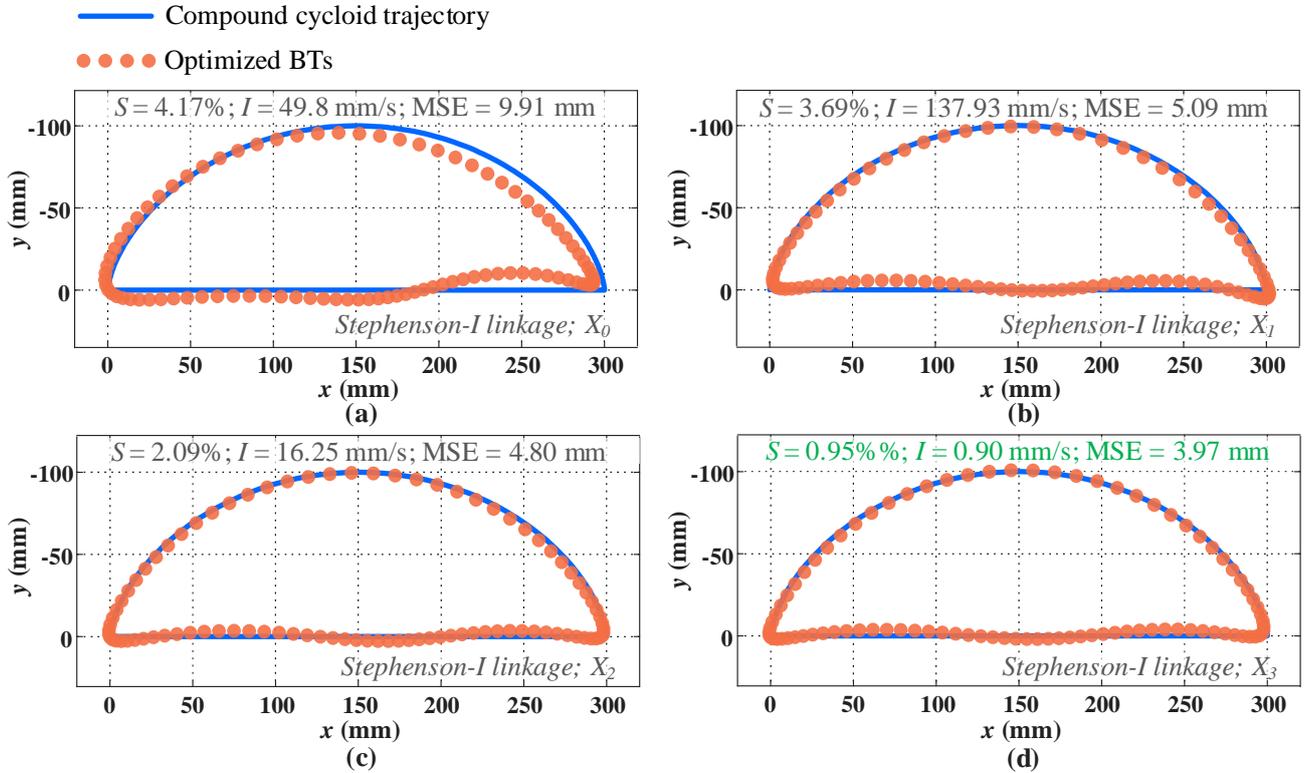

Fig. 6 Optimized trajectories with Stephenson-I linkage: (a) $X_0$, the trajectory synthesis result based on Fourier series, (b) $X_1$, the result of *Subtask 1*, (c) $X_2$, the result of *Subtask 2*, (d) $X_3$, the result of *Subtask 3*.

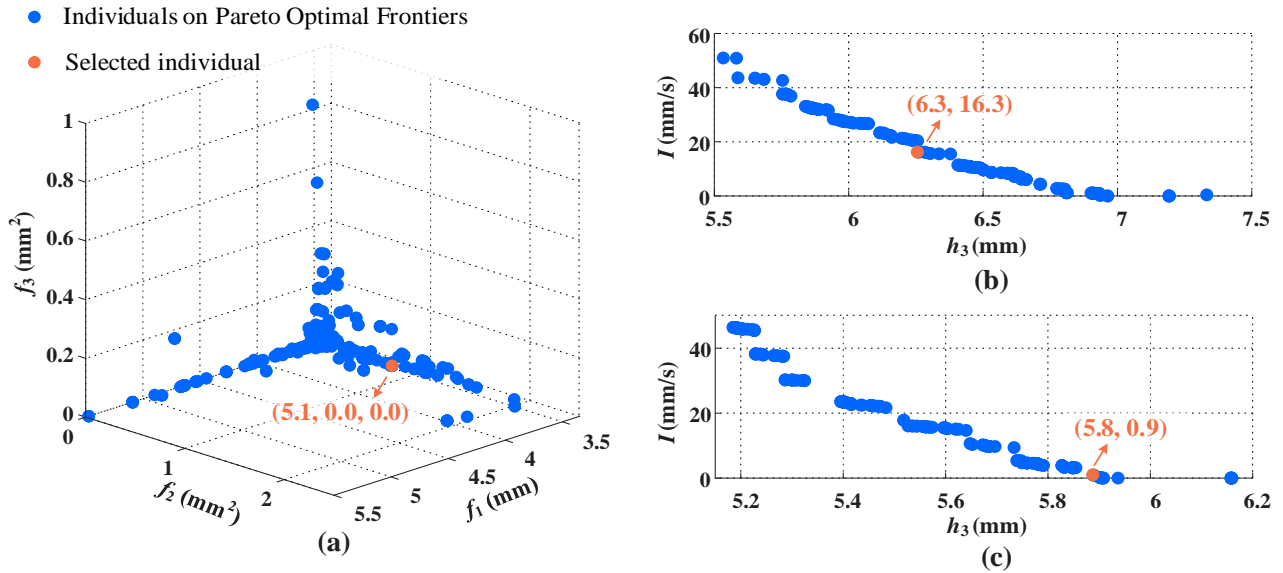

Fig. 7 Optimization terminates when individuals are distributed on a Pareto surface/curve: (a) *Subtask 1*, (b) *Subtask 2*, and (c) *Subtask 3*.

3.3 Simulation and prototype

To evaluate the design method outlined in section 3.2, we conducted a virtual prototype design and kinematic simulation. This involved using a pantograph linkage, specifically a parallelogram linkage, as detailed in references [53-55] and depicted in Fig. 8 (a). The linkage's length ratio (*JI*:*IL*:*LK* = 0.34:0.36:0.3) mirrors that of a cat's femur, tibia-fibula, and tarsometatarsus, resulting in a trajectory magnification factor of 1.88. We scaled the mechanism to ensure a BT length of 300 mm. Virtual prototypes of a quadruped module, are designed as shown in Fig. 8 (b). Using ADAMS, a quadruped module is simulated. The simulation results confirm the validity of Eq. 5, as the step length of the WT is 600 mm, twice the length of the BT stance phase. Fig. 8 (d) shows the trunk mass's vertical fluctuation during the swing phase, peaking at 5.8 mm, which aligns with the calculated value of $h_3$, validating the numerical calculation approach from BT to WT as per section 2.2. Additionally, Fig. 8 (e) shows the vertical velocity of the BT foot-end end movement, with a near-zero value at the landing point ($t = 0$). These results demonstrate that our design method successfully regenerates the shape and functions of compound cycloidal trajectory.

A single-legged prototype with Stephenson-I linkage was developed to evaluate the feasibility of the proposed design. A quadruped or multi-legged prototype was not pursued, as this presents an engineering issue rather than an academic one, with feasibility already established in prior studies [22-24]. The single-legged prototype is powered by a lithium battery, with motor control facilitated by a Bluetooth module. Further design specifics are omitted for brevity. The prototype's configurations at various crank positions are depicted in Fig. 9. Bench test experiments indicate that the prototype's actual trajectory closely matches the designed bench trajectory, thereby validating the effectiveness of the proposed approach for CLM design with trajectory shape constraints.

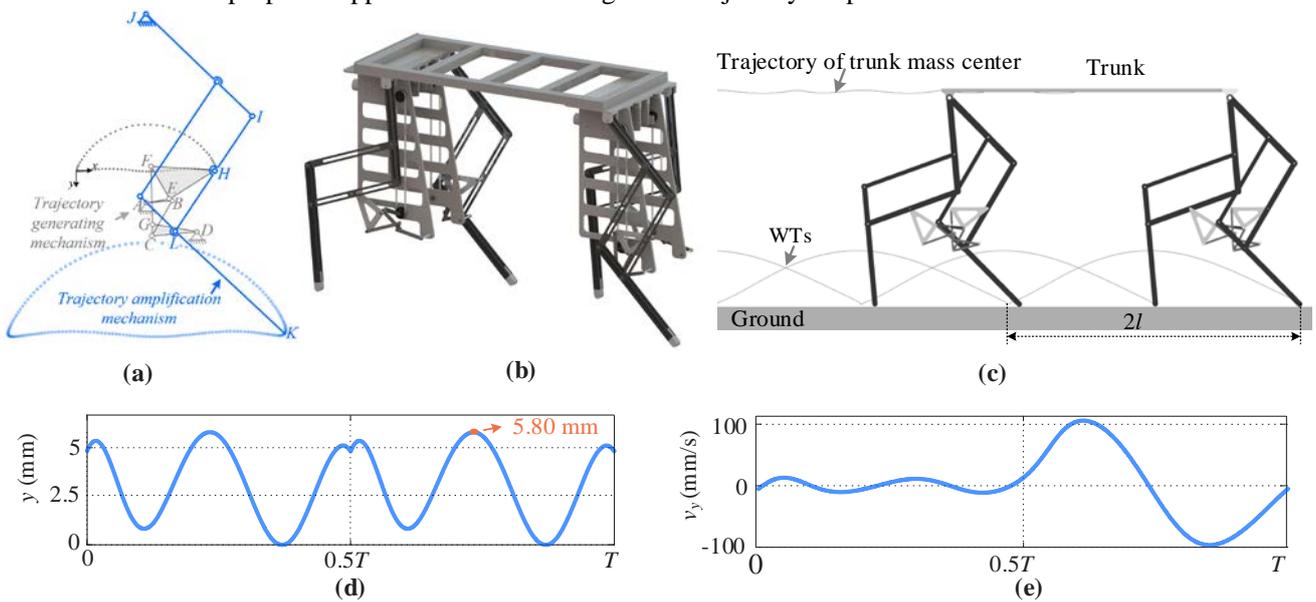

Fig. 8 Virtual prototype and simulation: (a) kinematic diagram, (b) virtual prototype of quadruped CLM module, (c) ADAMS simulation model, (d) vertical fluctuation of the trunk mass center, and (e) vertical velocity component of the foot endpoint of BT.

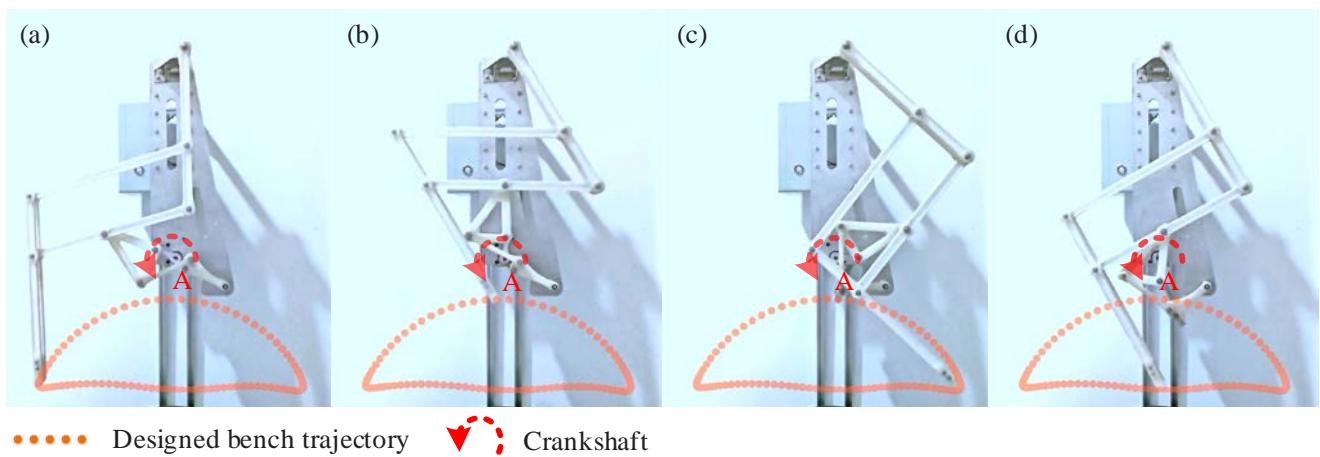

Fig. 9 Prototype experiment of CLM with the compound cycloid trajectory.

## 3.4 Validation and Comparison

### 3.4.1 Validation on other configurations

The above process was applied to the CLM design on Watt-I and Stephenson-III linkages to evaluate its generality. We did not discuss the Stephenson-II linkage because its first loop contains five links, which requires solving equations to compute the kinematics, thereby reducing computational efficiency. Regarding the Watt-II linkage, it can be considered a four-bar mechanism. The optimization results of Watt-I and Stephenson-III linkages are detailed in Appendix B-1 and Fig. 10. For the Watt-I linkage, the trajectory exhibits an MSE of 5.29 mm, an $S$ value of 0.75%, and an $I$ value of 14.3 mm/s. The Stephenson-III trajectory has an MSE of 7.12 mm, an $S$ value of 0.95%, and an $I$ value of 1.87 mm/s. These metrics indicate that both linkages maintain low fluctuations during the stance phase and ground impact, and the trajectory shapes are basically consistent with the planned compound cycloid trajectory. In summary, the hierarchical strategy is applicable to a variety of linkage configurations.

### 3.4.2 Validation on other optimization algorithms

To further validate the applicability of the hierarchical strategy across multiple algorithms, six state-of-the-art constrained multi-objective optimization algorithms published in 2024 (CMOEMT [56], CMOES [57], DRLOS-EMCMO [58], IMTCMO [59], MCCMO [60], and MOEA/D-CMT [61]) were incorporated. Each algorithm was executed at least three times, and the best-performing individuals were recorded. Fig. 11 shows the changes in the three metrics of the selected individual during the optimization process. Among all the algorithms, IMTCMO demonstrates superior performance compared to NSGA-II, while CMOEMT, DRLOS-EMCMO, and MOEA/D-CMT outperform NSGA-II in certain metrics. Further, Fig. 11 effectively illustrates the optimization direction of the three subtasks: The overall shape optimization (from $X_0$ to $X_1$) reduces the MSE but does not constrain $S$ and $I$, leading to an increase in these metrics. The trajectory straightness optimization (from $X_1$ to $X_2$) lowers $S$ while keeping MSE within an acceptable range. Finally, the local foot-ground impact optimization (from $X_2$ to $X_3$) significantly reduces $I$ values, while maintaining MSE and $S$ within reasonable limits. The individual values of $X_1$, $X_2$, and $X_3$ during the optimization process are shown in Appendix B-2 and B-3. In summary, the hierarchical strategy is applicable to a variety of existing multi-objective optimization algorithms.

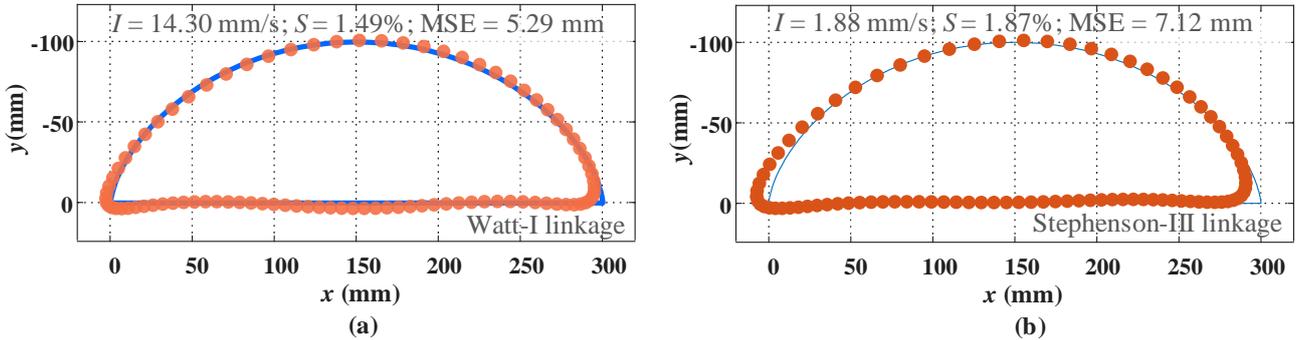

Fig. 10 Optimized results of other linkage configurations: (a) Watt-I linkage (b) Stephenson-III linkage.

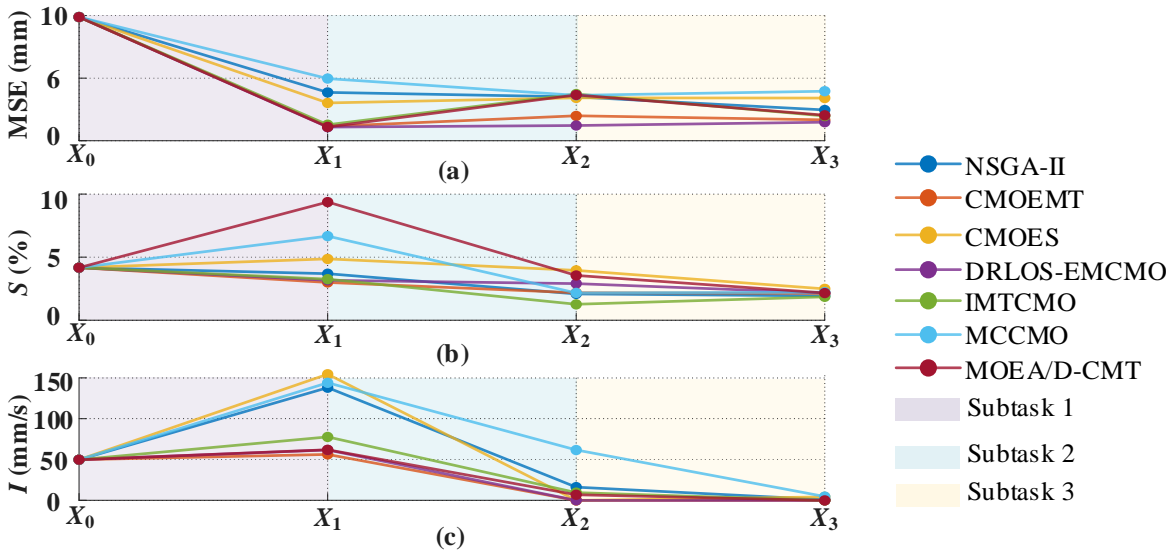

Fig.11 The evaluation values for three performance metrics corresponding to different algorithms

### 3.4.3 Comparison with single-level optimization

To illustrate the advantages, we compared hierarchical optimization with direct optimization (single-level optimization) using five objective functions. The same optimization algorithm and parameters were employed. Results are shown in Fig. 12. Fig. 12 (a) shows the objective space plot of all individuals, with all lying on the Pareto front. Figs. 12 (b-d) display the three individuals with the smallest values for $f_1$ (MSE), $f_4$ ($h_3$), and $f_5$ ($I$), respectively. Appendix B-4 presents the evaluation parameters for these individuals. The results indicate that direct optimization struggles to balance overall shape and local details. This is especially evident in the individual with the minimal $I$, whose MSE and S values are suboptimal. Although this individual achieves an $I$ value of almost 0, it shows no significant improvement over our result of 0.91 mm/s. In summary, the hierarchical strategy offers significant advantages over single-level optimization.

### 3.4.4 Comparison with other CLM design methods

The proposed method was compared with several common CLM design methods, include (1) the DP+SQP method: representing the BT using discrete points (DP) and optimizing the MSE function using the deterministic algorithm, Sequential Quadratic Programming (SQP); (2) the DP+GA method: representing the trajectory using DP and optimizing the MSE function with the stochastic algorithm, Genetic Algorithm (GA); (3) the DP+GA+SQP method: representing the trajectory using DP and sequentially optimizing the MSE function with GA followed by SQP; and (4) a combined optimization method based on Fourier descriptor and DP (this method corresponds to the results shown in Fig. 6a and has already been discussed in Section 3.2.3, so it will not be elaborated on here). Some studies have conducted similar work using Newton gradients or ant colony algorithms, but they are not fundamentally different from SQP or GA, so they will not be further elaborated. Each method is illustrated using the Stephenson-I mechanism, with the resulting optimized trajectories and parameters displayed in Fig. 13 and Appendix B-5. Our proposed method shows significant improvements in the three performances metrics: $I$, $S$, and MSE. In summary, the hierarchical strategy offers significant advantages over existing CLM design methods.

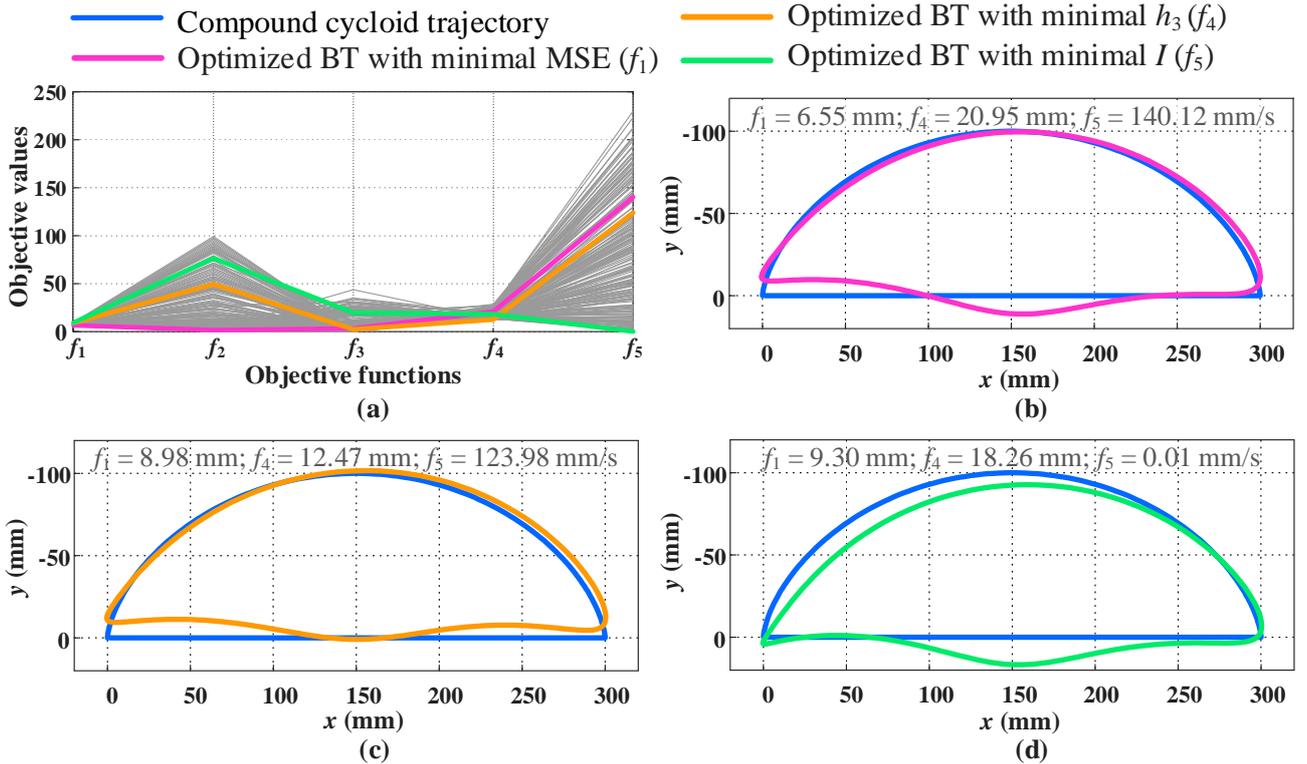

Fig.12 Results of the single-level optimization on five design objectives: (a) Objective space plot showing all 200 individuals converging to the Pareto-optimal front. Three individuals with the smallest values for $f_1$ (MSE), $f_4$ ($h_3$), and $f_5$ ($I$) are selected; (b) Optimized individual with minimal MSE, shown in pink; (c) optimized individual with minimal $h_3$, shown in orange; and (d) optimized individual with minimal $I$, shown in green.

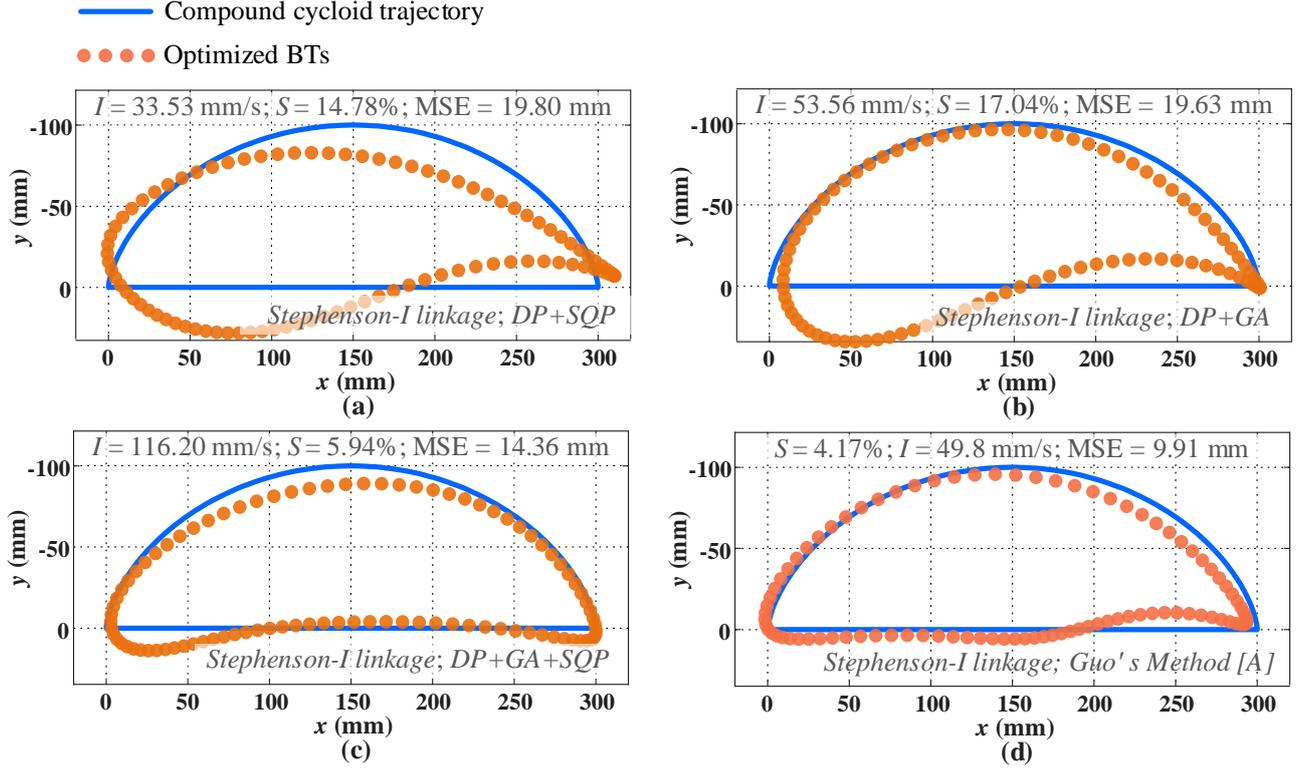

Fig. 13 Comparison with other optimal trajectory synthesis methods: (a) our method, (b) DP+SQP, (c) DP+GA, and (d) DP+GA+SQP.

## *4 CLM design without trajectory shape constraints*

This section presents the design of reconfigurable topology CLM (RTCLM) without trajectory shape constraints. A stepwise optimization process has been developed to ensure stability and minimize energy consumption in the *primary mode*, while also enhancing the obstacle-crossing height, mobility, and impact performance of the *auxiliary mode*. The design's feasibility is validated through simulations and a prototype, which substantiate its practicality. The efficacy of the strategy is further demonstrated through the validation of seven multi-objective optimization algorithms.

4.1 RTCLM analysis

An extra driving crank was added in a six-bar Watt-I linkage [24], creating an RTCLM, as shown in Fig. 12(a). This design allows it to switch smoothly between two topological structures: a six-bar (*primary mode*) configuration and a four-bar configuration (*auxiliary mode*). This switch is achieved by locking and activating joint motors, as shown in Fig. 12 (b) and (c). Define the switching state of two modes as follows: when links *AB* and *HF* are at specific angles (denoted as $\Phi_A$ and $\Phi_H$, respectively), the mechanism can switch to any mode. The two operating modes of this mechanism are explained as follows:

*Primary mode*: Joint *H* is locked, and the *DCFH* components form a distinct unit. The mechanism takes the form of a Watt six-bar linkage with crank AB, as shown in Fig. 14 (b)

*Auxiliary mode*: Joint Motor *A* remains locked, and the assembly *ABECHD* serves as a frame. The mechanism consists of a four-bar linkage with crank *HF*, as shown in Fig. 14 (c).

In a bipedal module, the two legs are categorized as left (*l*) and right (*r*) with a right superscript. To keep this RTCLM stable, the trajectories of the two legs must meet the following dimensional coupling constraints:

(1) In the *auxiliary mode*, the BTs exhibit no positional deviation in the *y*-direction. This is captured by $y_E(\Phi^l_A) = y_E(\Phi^r_A)$, where, $y_E$ denote the y-coordinate of point *E*, and $\Phi^l = \Phi^r + \pi$. The values of $\Phi^l_A$ and $\Phi^l_B$ are obtained through kinematics analysis.

(2) In the *switching state*, the BTs maintain a consistent shape. This implies that the vector **EH** for both the left and right legs remains identical:

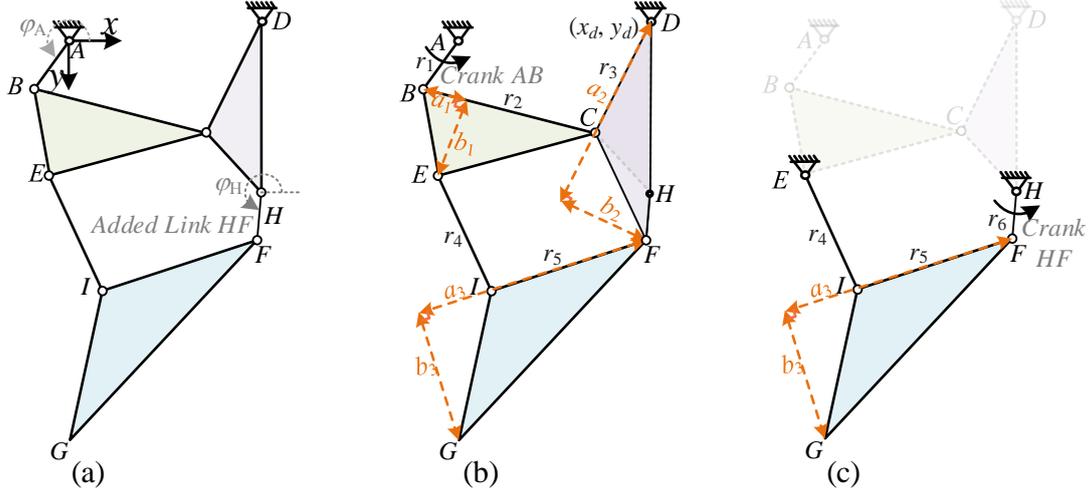

Fig. 14 Schematic diagram of mechanisms: (a) RTCLM with a seven-bar linkage, (b) the *primary mode*, and (c) the *auxiliary mode*.

$$\begin{cases} x_H^r(\Phi_A^l) = x_E^r(\Phi_A^l) + \Delta x_{EH} \\ x_H^l(\Phi_A^l) = x_H^l(\Phi_A^l) + \Delta x_{EH} \\ y_H(\Phi_A^l) = y_E(\Phi_A^l) + \Delta y_{EH} \end{cases} \quad (24)$$

where, **EH** = ($\Delta x_{EH}$, $\Delta y_{EH}$).

(3) In the *primary mode*, the virtual components *DCF* of both the left and right legs are congruent:

$$\begin{cases} (x_F^r - x_d)^2 + (y_F^r - y_d)^2 = (x_F^l - x_d)^2 + (y_F^l - y_d)^2 \\ (x_F^r - x_C^r)^2 + (y_F^r - y_C^r)^2 = (x_F^l - x_C^r)^2 + (y_F^l - y_C^r)^2 \end{cases} \quad (25)$$

where,

$$\begin{cases} x_F^r = x_H^r + r_6 \cos(\Phi_H^r) \\ y_F^r = y_H + r_6 \sin(\Phi_H^r) \\ x_F^l = x_H^l + r_6 \cos(\Phi_H^l) \\ y_F^l = y_H + r_6 \sin(\Phi_H^l) \end{cases} \quad (26)$$

By combining Eqs. (24-26), we get a system with three equations and four unknowns. $\Delta x_{EH}$ is chosen as the design variable, allowing to express $\Phi_H^l$, $\Phi_H^r$ and $\Delta y_{EH}$ in terms of $\Delta x_{EH}$. By solving Eqs. 26 and 27, the coordinates of points *F* and *H* can be deduced. Subsequently, the values of $a_2$, $b_2$, and the lengths of *DF* and *CH* can be calculated using the cosine theorem. In the absence of a predefined requirement for relative trajectory positioning, we assume that $x_a = y_a = 0$. Thus, the mechanism's design parameters are set as $X = [r_1\ r_2\ r_3\ r_4\ r_5\ r_6\ x_d\ y_d\ a_1\ b_1\ a_3\ b_3\ \Delta x_{EH}]$, with the remaining parameters derived from these dimensional coupling relationships.

4.2 Performance design based on hierarchical strategy

4.2.1 Objective functions

The design parameters are *X*. The objective functions for the design optimization are

$$\text{Min.} \begin{cases} f_1 = (h_4 - h_{4,o})^2 + g(X) \\ f_2 = (h_6 - h_{6,o})^2 + g(X) \end{cases} \quad (27)$$

where, the penalty function $g(X)$ is designed to guarantee the uninterrupted progression of the program, even in scenarios where optimized individuals demonstrate linkage defects or if the solutions to the equations are flawed. In the absence of this function, these circumstances could potentially disrupt the optimization process or result in incoherent motion trajectories for the left and right legs. $h_{6,t}$ and $h_{4,t}$ denote the target height performance of the *primary* and *auxiliary modes*, respectively.

4.2.2 Constrains
(1) The search ranges $X_{lb}$ and $X_{ub}$ are shown in Appendix C-1.
$$X_{lb} \leqslant X \leqslant X_{ub} \tag{28}$$
(2) Branch and order defects do not occur when the Assur Group Method is used to solve kinematics. Loop defects will lead to imaginary numbers, which are bounded by the penalty function $g(X)$. Crank defects are avoided by:
$$\text{S.t.} \begin{cases} r_1 + 2\max(r_2, r_3, (x_d^2 + y_d^2)^{1/2}) < r_2 + r_3 + (x_d^2 + y_d^2)^{1/2} \\ r_6 + 2\max(r_4, r_5, (\Delta x_{EH}^2 + \Delta y_{EH}^2)^{1/2}) < r_4 + r_5 + (\Delta x_{EH}^2 + \Delta y_{EH}^2)^{1/2} \end{cases} \tag{29}$$
(3) To prevent retrograde motion during the movement of CLM, it is imperative to adhere to a specific order of $t_1$, $t_2$, $t_5$, and $t_4$. Additionally, the landing and take-off angles need to exceed a minimum threshold of 15 degrees:
$$\text{S.t.} \begin{cases} (t_2 < t_5 < t_4 < t_1) \ or \ (t_5 < t_4 < t_1 < t_2) \ or \ (t_4 < t_1 < t_2 < t_5) \ or \ (t_1 < t_2 < t_5 < t_4) \\ \theta_1 > 15 \\ \theta_2 > 15 \end{cases} \tag{30}$$
(4) In the *auxiliary mode*, the average crossing height of a single trajectory shall not be less than one fourth of the maximum crossing height; The probability of non-contact crossing of a block obstacle with a length of 25 mm and a height of 25 mm on a quadruped trajectory is greater than 20%
$$\text{S.t.} \begin{cases} \bar{h} > h_m / 4 \\ \psi_2(25, 25) > 20\% \end{cases} \tag{31}$$

4.2.3 Hierarchical strategy and results

The stepwise solution strategy consists of two subtasks. In the first subtask, the *primary mode* motion performance is optimized, selecting individuals with small $f_2$ optimization deviations. In the second subtask, the individuals from the first phase are used as constraints, and the motion performance in the *auxiliary mode* is optimized by selecting individuals with small $f_1$ deviations while adding the obstacle crossing probability constraint from Eq. 31. Using NSGA-II as an example, the population size was set to 50, with both the crossover and mutation fractions set to 2/13 (based on sensitivity analysis). The target height performance for *primary mode* $h_{6,\,t}$ was set to 50 mm, and the target height performance for the *auxiliary mode* $h_{4,\,t}$ was set to 220 mm (case 1), 240 mm (case 2), 260 mm (case 3), 280 mm (case 4), and 300 mm (case 5), respectively.

Fig. 15 shows the distribution of individuals for each case, from which one individual on the Pareto-optimal frontier was selected. The corresponding BTs and WTs are shown in Fig. 16, with further details in Appendix C-1. The results show that the synthesized trajectory height of the RTCLM in *primary mode* remains within a narrow range of 50 ± 0.32 mm, while in auxiliary mode, the deviation is under 0.5 mm from the target height, achieving precise obstacle-crossing design. The five cases show no significant nondominated relationships between the two objective functions, meaning a clear Pareto optimal curve does not emerge. Limiting iterations to a maximum of 50 accelerates the solution process. To balance design time with accuracy, optimization stops automatically when individual objective values fall below the threshold [0.5, 0.5]. This typically occurs within 40 generations, keeping computational time to around 40 minutes per run (using MATLAB on an Intel 12700KF single-core processor without GPU acceleration).

4.3 Simulation and prototype

In the design exemplified by case 1, a prototype of an eight-legged robot has been designed, as shown in Fig. 17 (a). In the *primary mode*, a motor drives the crank AB via a worm-gear mechanism. In the *auxiliary mode*, a separate motor drives the crank HF through a worm-gear and wheel belt drive. The trajectory simulations for the BT and WT tests were performed in Adams, as shown in Figs. 17 (b-e). The congruence between the simulated and calculated trajectories confirms the feasibility of the design approach.

A single-legged prototype was developed to evaluate the feasibility of the proposed design, as illustrated in Fig. 18. Crankshaft A was driven by a motor-worm gear drive system, while crankshaft B was driven by a motor-worm gear-belt drive system. Figs. 18 (a) and (b) depict the prototype's actual trajectory under *primary* and *auxiliary modes*, respectively. The results demonstrate that the prototype's actual trajectories closely aligns with the designed trajectories in both modes, thereby validating the effectiveness of the trajectory performance-driven optimal dimensional synthesis strategy for CLM design without trajectory shape constraints.

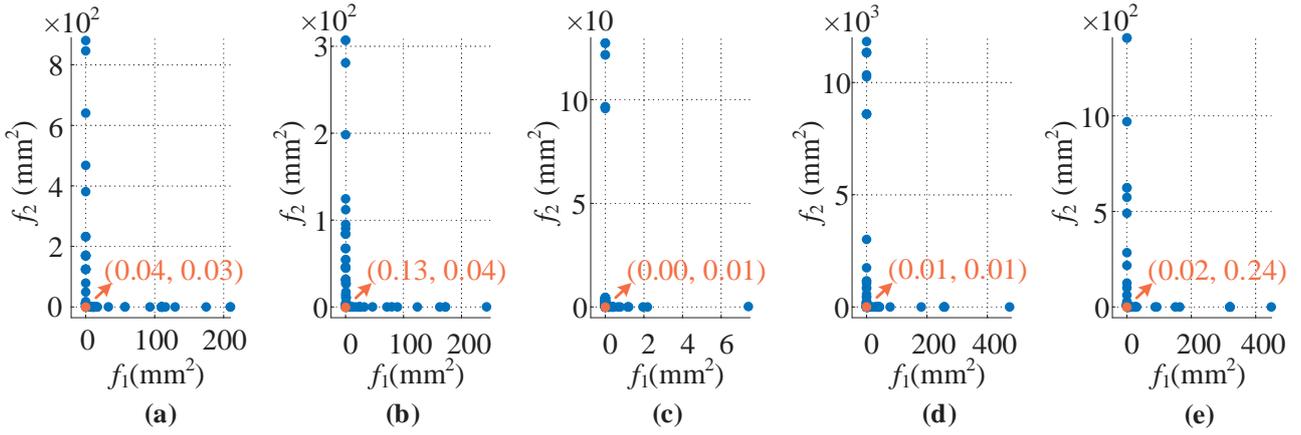

Fig. 15 Individual distribution at the end of optimization: (a) case 1, (b) case 2, (c) case 3, (d) case 4, (e) case 5.

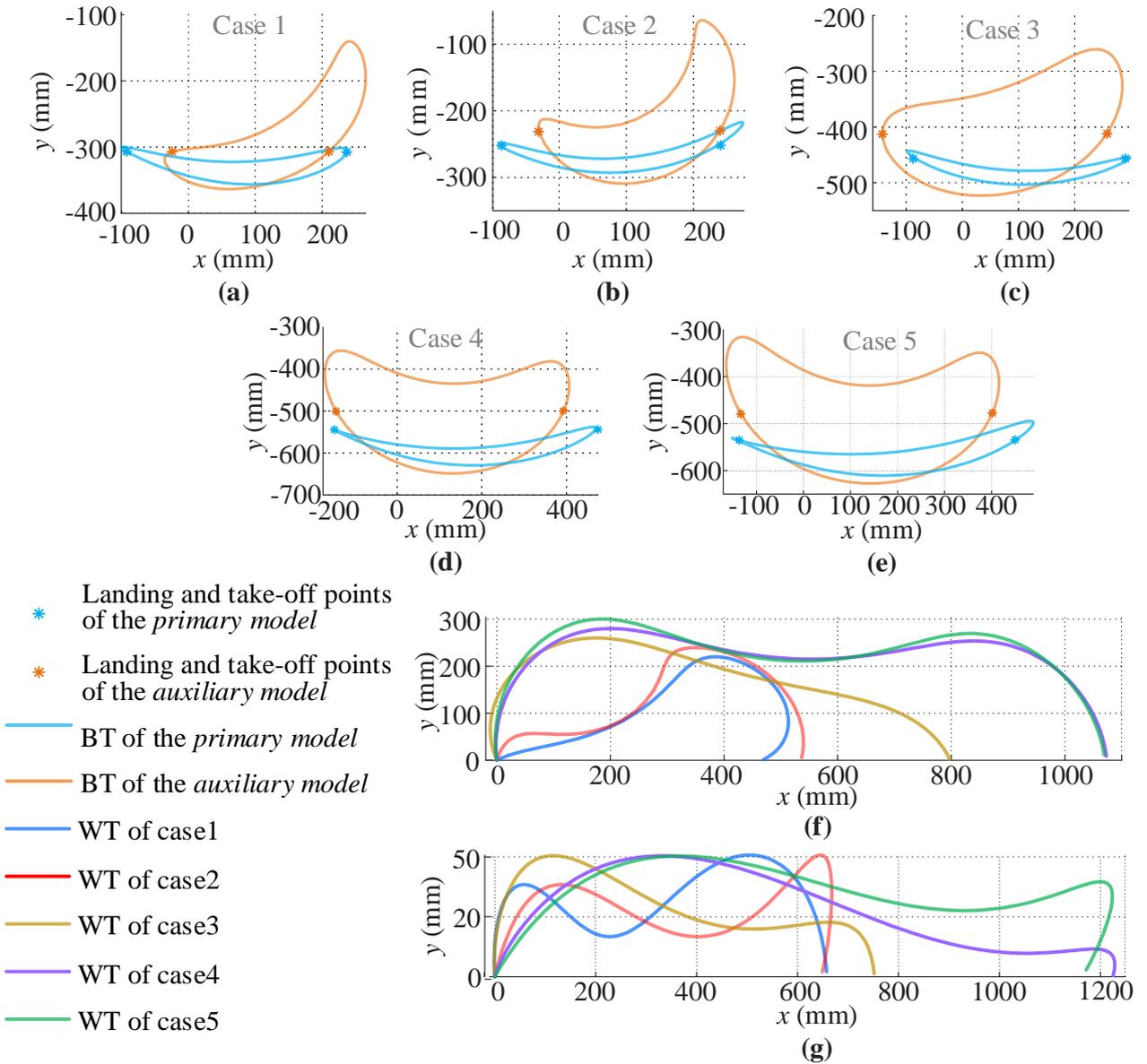

Fig. 16 BT and WT of the selected individuals: (a-e) BTs, (f) WTs in the *auxiliary mode*, and (g) WTs in the *primary mode*.

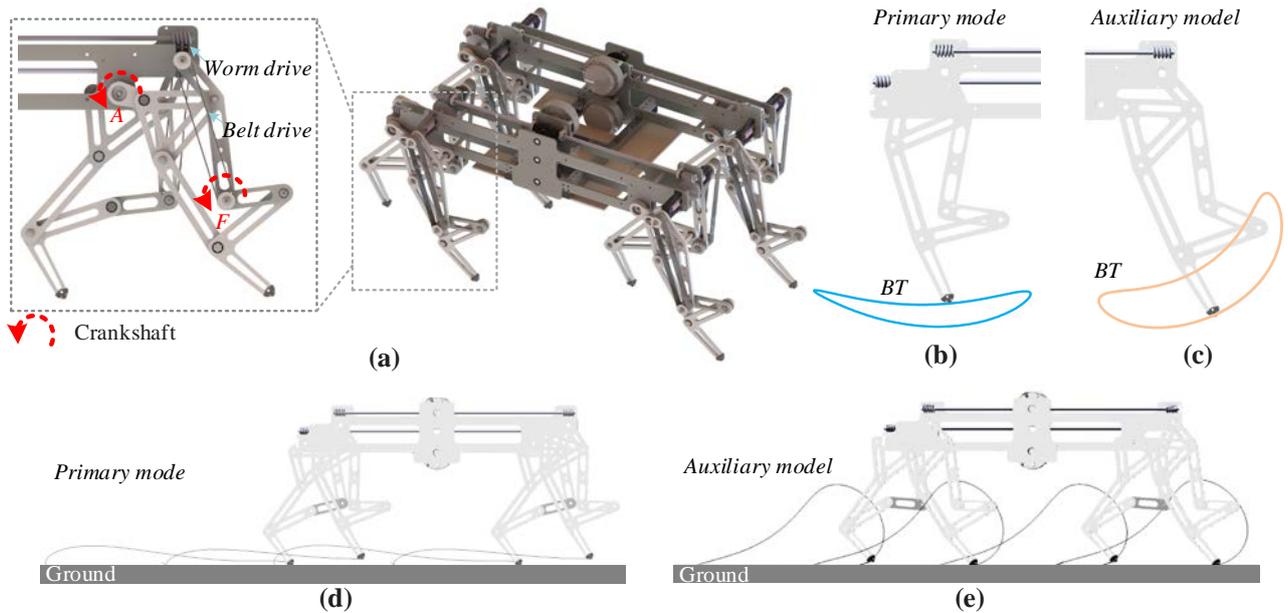

Fig. 17 Solid model and simulations: (a) virtual solid prototype, (b) simulation of BT in the *primary mode*, (c) simulation of BT in the *auxiliary mode*, (d) simulation of WT in the *primary mode*, and (e) simulation of WT in the *auxiliary mode*.

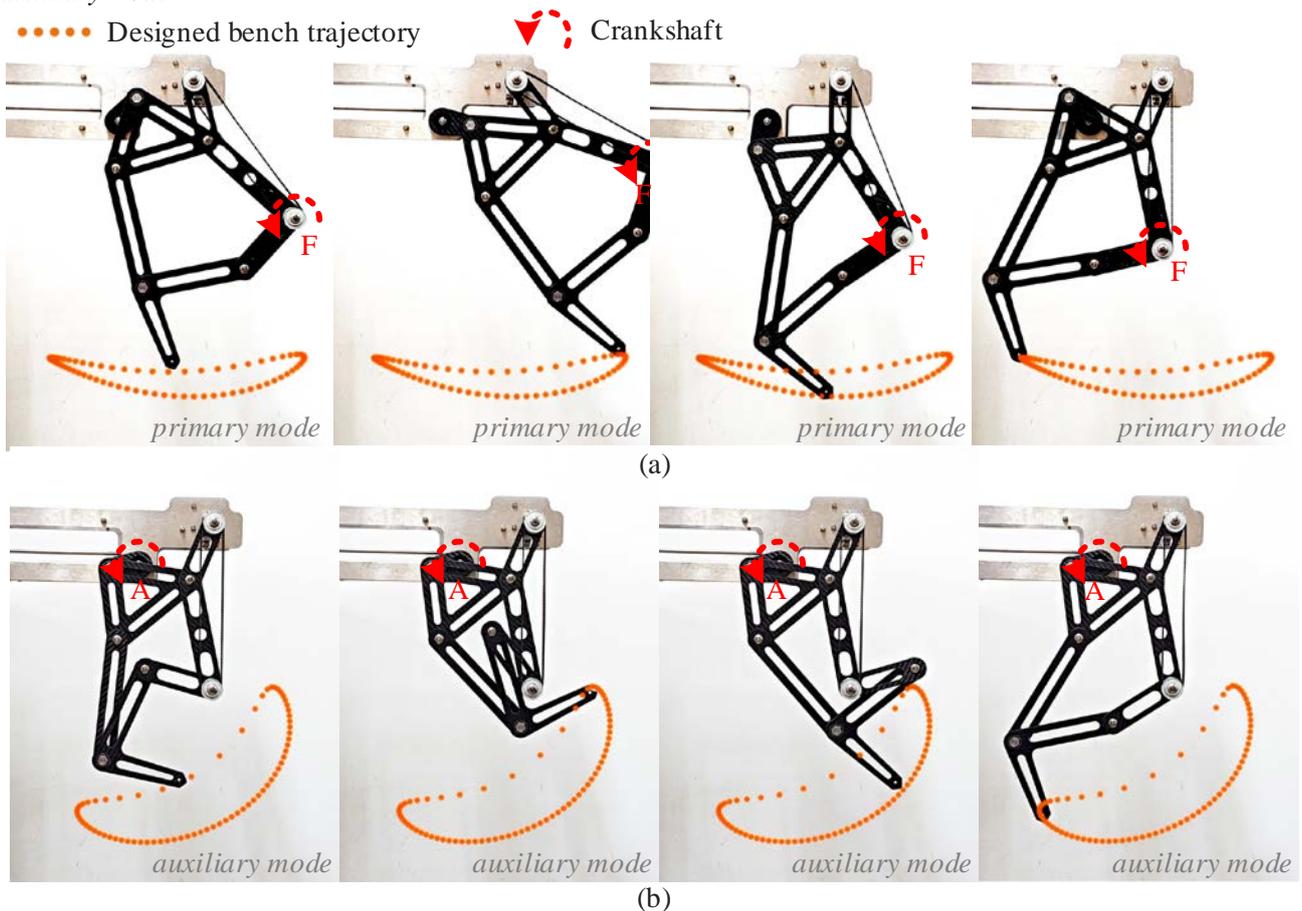

Fig. 18 Prototype experiments of the design result of case 1: (a) the *primary mode* and (b) the *auxiliary mode*.

4.4 Validation and Comparison

Similar to Section 3.4.2, six algorithms were used for validation, with all parameters consistent with those in Section 4.2. The results were obtained by solving the problem at least three times and selecting the optimal outcome. The algorithm performance was evaluated using both absolute and relative deviations. Absolute deviation on 5 cases is defined as $\Sigma(|h_{4,o} - h_{4,t}| + |h_{6,o} - h_{6,t}|)/5$, and relative deviations are defined as $\Sigma|h_{6,o} - h_{6,t}|/(5 \times h_{6,t})$ for the *primary mode* and $\Sigma|h_{4,o} - h_{4,t}|/(5 \times h_{4,t})$ for the *auxiliary mode*, where $h_{6,o}$ and $h_{4,o}$ *denote the* optimized parameters.

The deviation results are shown in Fig. 19. The values of the selected individual by different algorithms for the five cases are shown in Appendix C-2 to Appendix C-6. The results indicate that the absolute and relative deviations of the CMOEMT, DRLOS-EMCMO, and MOEA/D-CMT algorithms are better than those of NSGA-II, with smaller standard deviations. IMTCMO exhibits performance similar to NSGA-II. On the other hand, CMOES and MCCMO have higher absolute and relative deviations, as well as larger standard deviations compared to NSGA-II. Nevertheless, even the worst-performing algorithm has an average absolute deviation not exceeding 1 mm and an average relative deviation not exceeding 7‰, suggesting that most of the algorithms meet the required performance criteria for practical applications.

This section introduces an RTCLM capable of switching between 4-bar and 6-bar configurations to validating the effectiveness of the trajectory performance-driven hierarchical strategy for CLM design. Unlike previous reconfigurable CLM studies, RTCLMs introduce unique challenges, such as the inability to linearly adjust rod sizes and more stringent coupling constraints with fewer design parameters. To our knowledge, this is the first introduction of RTCLM that successfully achieves precise control over obstacle-crossing performance parameters.

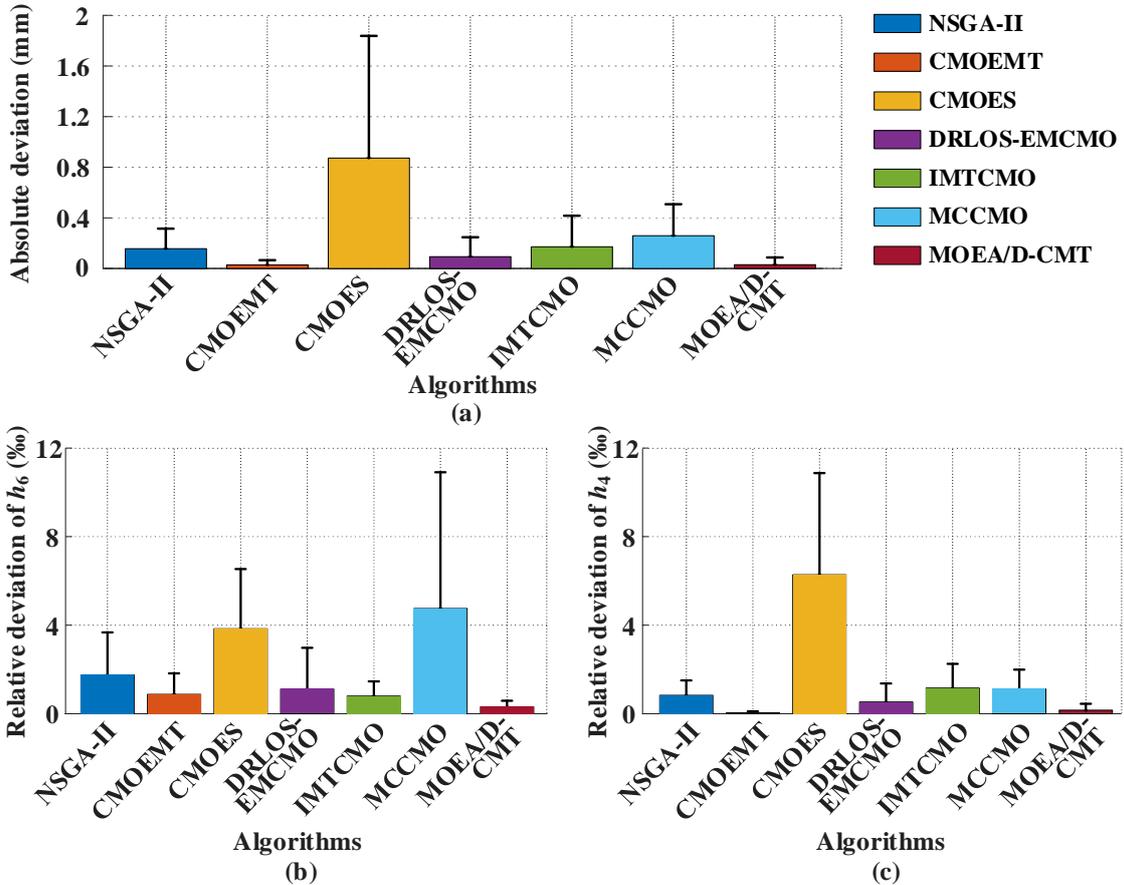

Fig. 19 Optimization deviations of different algorithms: (a) absolute deviations, (b) relative deviations of $h_4$, and (c) relative deviations of $h_6$.

## 5 Conclusion

This study developed a hierarchical multi-objective optimization strategy for the precise performance design of CLMs, both with and without trajectory shape constraints. The effectiveness of the hierarchical strategy fundamentally relies on the numerical performance-trajectory mapping we established, which classifies performance parameters into weakly coupled categories. Leveraging the characteristics of trajectory shapes and combining them with the hierarchical strategy, we successfully achieved the first-ever implementation of a compound cycloidal trajectory on a CLM, as well as the precise performance design for a reconfigurable CLM. This is a significant step forward in the development of high-performance closed-chain robots. A series of validations and comparisons were conducted. The results confirm the method's effectiveness across various configurations and algorithms, demonstrate superior performance compared to existing CLM design methods, and verify the capacity of the realized CLMs to be successfully replicated in the prototype. Future work will focus on integrating the designed CLMs with environmental perception, flexible motor control, and behavioral decision-making to develop high-performance closed-chain legged robots capable of precise, autonomous operation in complex environments.

There are some limitations: In high-dimensional objective spaces, knee-point identification becomes less precise as the geometric distinctiveness of Pareto front features diminishes. Additionally, the reliance on domain expertise for manual selection may somewhat limit its general applicability.

## Acknowledgements

This work was supported by the National Natural Science Foundation of China [Grant No. 52205007, 51175031].

## Nomenclature

| | |
|---|---|
| $C_b$, $x_b$, $y_b$ | Bench trajectory |
| $C_m$, $x_m$, $y_m$ | Walking trajectory |
| $N$ | Number of sampling points |
| $T$ | Period |
| $C^K$, $x^K$, $y^K$ | Trajectory of leg K |
| $^K C$ | Trajectories in {K} coordinate system |
| $l_s$ | Length of stance phase |
| $h_s$ | Height of stance phase |
| $l_s$ | Length of walking trajectory |
| $S$ | Straightness of stance phase |
| $I$ | Vertical velocity of landing point |
| $\theta$ | Angle of landing and takeoff points |
| $h_m$ | Maximum obstacle crossing height |
| $\bar{h}$ | Average obstacle crossing height |
| $\psi$ | Obstacle crossing probability |
| $\Phi$ | Link angle in switching state |
| $X$ | Design variables |

## Abbreviations

| | |
|---|---|
| BT | Bench trajectory |
| WT | Walking trajectory |
| CLM | Close-chain legged mechanism |
| RTCLM | Reconfigurable topological close-chain legged mechanism |

# Appendix

Appendix A

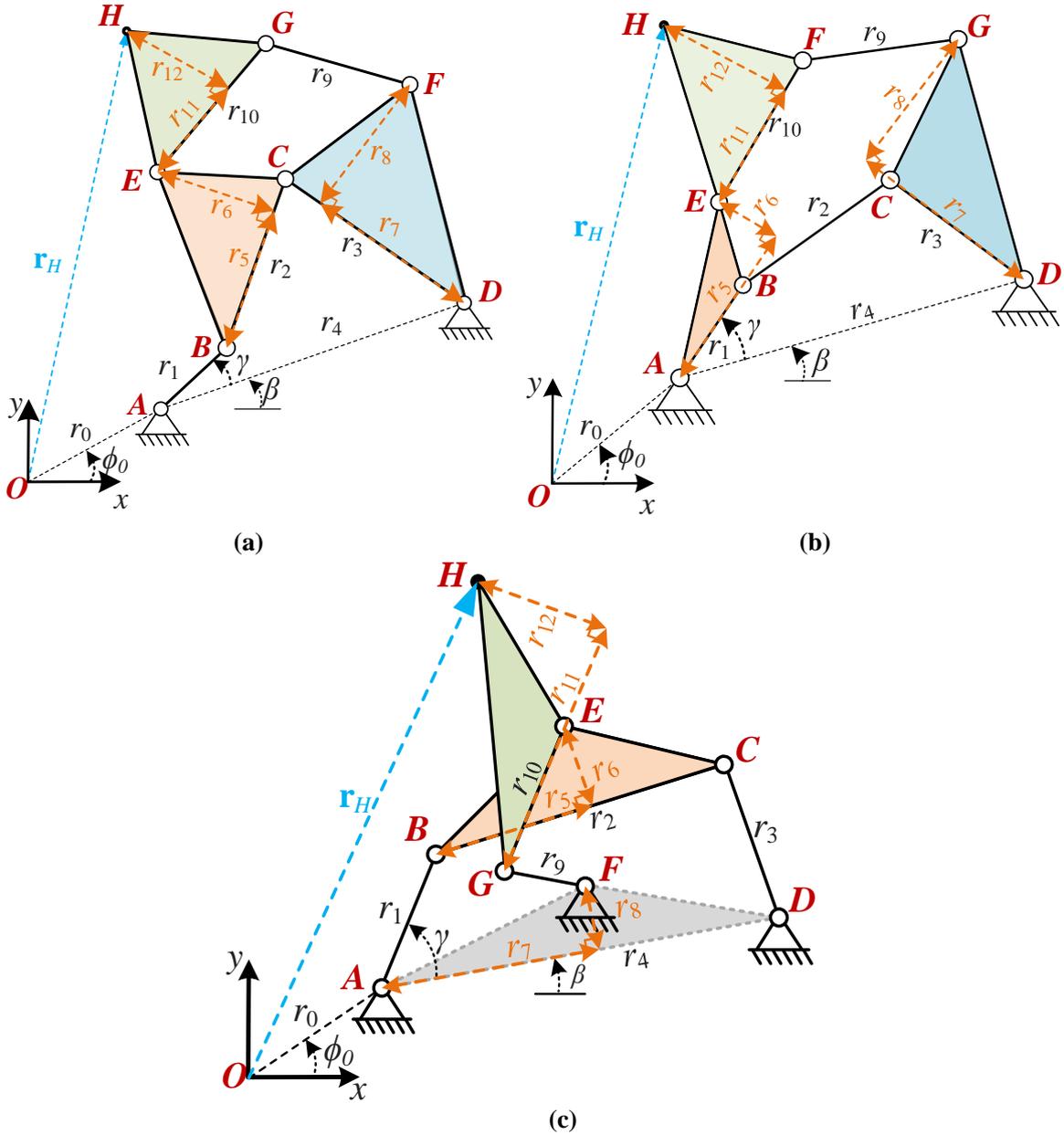

Appendix A: Schematic diagrams of linkage parameters: (a) Watt-I linkage, (b) Stephenson-I linkage, and (c) Stephenson-III linkage.

Appendix B

Appendix B-1: Optimization results and performance of different linkage configurations in Section 3.

| Results | Stephenson-I | | | | Watt-I | Stephenson-III |
|---|---|---|---|---|---|---|
| | $X_0$ | $X_1$ | $X_2$ | $X_3$ | | |
| *Time*(min) | 2.18 | 0.93 | 1.08 | 3.50 | 5.25 | 6.86 |
| S (%) | 4.17 | 3.69 | 2.09 | 1.93 | 0.75 | 0.95 |
| I (mm/s) | 49.8 | 137.93 | 16.25 | 0.91 | 14.31 | 1.87 |
| MSE (mm) | 9.91 | 5.09 | 4.80 | 3.97 | 5.29 | 7.12 |
| b (rad) | 0.31 | 0.49 | 0.48 | 0.52 | 6.10 | 1.12 |
| g (rad) | 3.14 | 3.22 | 3.52 | 3.26 | 2.46 | 3.50 |
| $x_a$ (mm) | 137.85 | 147.39 | 148.49 | 150.57 | 64.59 | 57.03 |
| $y_a$ (mm) | 84.31 | 67.92 | 68.98 | 66.12 | 148.82 | -151.91 |
| $r_1$ /mm | 49.63 | 56.55 | 55.47 | 56.20 | 90.88 | 82.02 |
| $r_2$ /mm | 78.12 | 90.12 | 85.14 | 88.15 | 177.77 | 371.54 |
| $r_3$ /mm | 87.94 | 99.70 | 102.75 | 99.62 | 180.98 | 226.37 |
| $r_4$ /mm | 102.20 | 125.27 | 126.49 | 128.33 | 191.94 | 259.41 |
| $r_5$ /mm | 50.25 | 49.75 | 50.62 | 51.20 | 214.01 | 384.43 |
| $r_6$ /mm | -6.13 | -4.13 | -5.88 | -4.21 | -177.72 | -324.09 |
| $r_7$ /mm | 86.79 | 96.13 | 100.09 | 96.50 | 62.41 | 343.95 |
| $r_8$ /mm | -21.20 | -32.06 | -25.46 | -25.13 | -50.08 | -22.07 |
| $r_9$ /mm | 142.52 | 132.34 | 129.48 | 127.65 | 200.26 | 395.95 |
| $r_{10}$ /mm | 101.39 | 93.38 | 88.80 | 83.21 | 209.76 | 361.26 |
| $r_{11}$ /mm | 4.96 | 2.23 | 6.64 | 8.09 | -239.21 | 411.72 |
| $r_{12}$ /mm | 129.56 | 117.97 | 117.53 | 115.54 | 29.84 | -355.19 |

Appendix B-2: Optimization results and performance of different algorithms in Section 3.

| Results | CMOEMT | | | CMOES | | | DRLOS-EMCMO | | |
|---|---|---|---|---|---|---|---|---|---|
| | $X_1$ | $X_2$ | $X_3$ | $X_1$ | $X_2$ | $X_3$ | $X_1$ | $X_2$ | $X_3$ |
| MSE (mm) | 2.91 | 3.59 | 3.33 | 4.42 | 4.74 | 4.73 | 2.87 | 2.98 | 3.19 |
| $S$ (%) | 3.00 | 2.18 | 2.13 | 4.86 | 3.96 | 2.49 | 3.17 | 2.91 | 2.16 |
| $I$ (mm/s) | 56.34 | 0.01 | 0.00 | 154.46 | 0.01 | 4.27 | 61.70 | 0.00 | 0.02 |
| $b$ (rad) | 0.49 | 0.51 | 0.51 | 0.76 | 0.70 | 0.65 | 0.49 | 0.48 | 0.47 |
| $g$ (rad) | 3.25 | 3.26 | 3.27 | 3.40 | 3.38 | 3.36 | 3.24 | 3.26 | 3.27 |
| $x_a$ (mm) | 150.14 | 150.67 | 151.13 | 152.05 | 153.95 | 153.46 | 149.97 | 150.36 | 150.70 |
| $y_a$ (mm) | 68.86 | 69.09 | 67.63 | 75.49 | 66.97 | 69.78 | 68.46 | 68.58 | 68.69 |
| $r_1$ /mm | 60.58 | 57.68 | 58.99 | 52.48 | 54.82 | 53.13 | 63.52 | 64.39 | 63.37 |
| $r_2$ /mm | 92.40 | 91.28 | 93.68 | 87.07 | 82.93 | 81.97 | 97.73 | 99.35 | 97.87 |
| $r_3$ /mm | 115.97 | 112.21 | 115.68 | 82.82 | 83.76 | 86.24 | 112.53 | 122.86 | 127.68 |
| $r_4$ /mm | 142.60 | 139.70 | 145.63 | 113.72 | 106.93 | 108.00 | 142.54 | 152.16 | 157.63 |
| $r_5$ /mm | 51.98 | 50.21 | 50.75 | 50.39 | 52.56 | 51.13 | 52.20 | 51.64 | 50.43 |
| $r_6$ /mm | -5.62 | -4.77 | -4.88 | -5.64 | -7.66 | -7.34 | -5.17 | -5.70 | -5.36 |
| $r_7$ /mm | 110.99 | 107.98 | 108.79 | 90.00 | 92.07 | 91.72 | 110.78 | 121.16 | 123.83 |
| $r_8$ /mm | -24.73 | -27.45 | -26.68 | -28.88 | -28.73 | -28.20 | -30.88 | -33.24 | -31.53 |
| $r_9$ /mm | 148.39 | 143.90 | 147.84 | 152.13 | 145.33 | 148.27 | 183.93 | 185.38 | 183.13 |
| $r_{10}$ /mm | 97.31 | 97.56 | 97.18 | 127.79 | 122.73 | 125.45 | 144.63 | 147.87 | 143.69 |
| $r_{11}$ /mm | 24.13 | 21.50 | 21.82 | 9.26 | 13.74 | 11.63 | 0.13 | 1.91 | 1.75 |
| $r_{12}$ /mm | 114.55 | 117.14 | 115.20 | 124.00 | 114.11 | 117.17 | 114.57 | 117.22 | 118.23 |

Appendix B-3: Optimization results and performance of different algorithms in Section 3.

| Results | IMTCMO | | | MCCMO | | | MOEA/D-CMT | | |
|---|---|---|---|---|---|---|---|---|---|
| | $X_1$ | $X_2$ | $X_3$ | $X_1$ | $X_2$ | $X_3$ | $X_1$ | $X_2$ | $X_3$ |
| MSE (mm) | 3.03 | 4.99 | 3.58 | 5.98 | 4.91 | 5.17 | 2.887 | 4.93 | 3.63 |
| $S$ (%) | 3.26 | 1.26 | 1.86 | 6.67 | 2.17 | 2.23 | 9.37 | 3.55 | 2.16 |
| $I$ (mm/s) | 77.65 | 9.51 | 0.00 | 143.95 | 61.65 | 4.64 | 61.91 | 6.88 | 0.01 |
| $b$ (rad) | 0.55 | 2346 | 047 | 1.03 | 0.51 | 0.59 | 0.49 | 0.49 | 0.48 |
| $g$ (rad) | 3.30 | 3.20 | 3.24 | 3.66 | 3.26 | 3.30 | 3.24 | 3.19 | 3.21 |
| $x_a$ (mm) | 150.94 | 151.98 | 150.89 | 155.50 | 150.67 | 151.54 | 150.06 | 151.70 | 151.24 |
| $y_a$ (mm) | 71.05 | 74.15 | 70.00 | 77.07 | 69.08 | 78.10 | 68.38 | 68.56 | 67.09 |
| $r_1$ /mm | 60.41 | 55.28 | 58.90 | 51.34 | 57.68 | 49.57 | 62.89 | 51.99 | 58.43 |
| $r_2$ /mm | 94.27 | 95.46 | 98.79 | 81.42 | 91.27 | 72.49 | 95.95 | 99.08 | 102.57 |
| $r_3$ /mm | 113.28 | 122.02 | 126.90 | 74.38 | 112.21 | 86.15 | 113.08 | 98.69 | 103.58 |
| $r_4$ /mm | 142.11 | 158.68 | 162.68 | 98.56 | 139.70 | 104.08 | 142.64 | 141.71 | 145.23 |
| $r_5$ /mm | 5177 | 48.94 | 50.05 | 50.35 | 50.21 | 48.04 | 52.17 | 49.78 | 50.85 |
| $r_6$ /mm | -5.96 | -1.63 | -3.64 | -13.85 | -4.77 | -7.13 | -5.23 | -0.33 | -1.72 |
| $r_7$ /mm | 109.09 | 118.49 | 119.96 | 89.46 | 107.98 | 95.27 | 110.60 | 96.35 | 99.99 |

| | | | | | | | | |
|---|---|---|---|---|---|---|---|---|
| $r_8$ /mm | -26.63 | -25.10 | -26.51 | -25.14 | -27.45 | -19.07 | -30.14 | -30.69 | -33.05 |
| $r_9$ /mm | 153.03 | 153.24 | 157.22 | 115.72 | 143.90 | 123.52 | 163.10 | 183.03 | 181.53 |
| $r_{10}$ /mm | 104.49 | 96.55 | 98.18 | 117.23 | 97.56 | 104.07 | 122.17 | 133.67 | 130.97 |
| $r_{11}$ /mm | 23.91 | 26.59 | 25.07 | 0.01 | 21.49 | 3.38 | 0.70 | 3.24 | 4.79 |
| $r_{12}$ /mm | 116.64 | 120.07 | 117.00 | 125.56 | 117.14 | 126.35 | 116.50 | 11.45 | 116.48 |

Appendix B-4 Comparison of our method with single-level optimization in Section 3.

| Results | Hierarchical method of ours | Direct multi-objective optimization | | |
|---|---|---|---|---|
| | | with smallest MSE | with smallest $h_3$ | with smallest $I$ |
| MSE (mm) | 3.97 | 6.55 | 8.98 | 9.30 |
| $S$ (%) | 1.93 | 6.98 | 4.16 | 6.09 |
| $I$ (mm/s) | 0.91 | 140.12 | 123.98 | **0.01** |
| $b$ (rad) | 0.52 | 0.81 | 0.81 | 0.81 |
| $g$ (rad) | 3.26 | 3.43 | 3.43 | 3.43 |
| $x_a$ (mm) | 150.57 | 158.46 | 162.69 | 161.23 |
| $y_a$ (mm) | 66.12 | 82.04 | 81.18 | 84.59 |
| $r_1$ /mm | 56.20 | 62.58 | 63.91 | 63.98 |
| $r_2$ /mm | 88.15 | 104.26 | 103.62 | 104.18 |
| $r_3$ /mm | 99.62 | 94.74 | 98.07 | 94.42 |
| $r_4$ /mm | 128.33 | 132.42 | 132.42 | 132.42 |
| $r_5$ /mm | 51.20 | 55.55 | 53.09 | 56.52 |
| $r_6$ /mm | -4.21 | -8.38 | -8.10 | -8.14 |
| $r_7$ /mm | 96.50 | 94.68 | 88.71 | 94.68 |
| $r_8$ /mm | -25.13 | -28.78 | -28.68 | -29.30 |
| $r_9$ /mm | 127.65 | 122.79 | 123.35 | 122.79 |
| $r_{10}$ /mm | 83.21 | 82.89 | 82.85 | 82.82 |
| $r_{11}$ /mm | 8.09 | 4.46 | 8.39 | 8.50 |
| $r_{12}$ /mm | 115.54 | 125.41 | 128.43 | 124.69 |

Appendix B-5: Comparison with Other Trajectory Synthesis Methods in Section 3.

| | $l_b$ | $u_b$ | DP + SQP | DP + GA | DP + GA + SQP | FS and DP-based method [9] | Our method |
|---|---|---|---|---|---|---|---|
| Time(min) | / | / | $1.6\times10^{-4}$ | 0.32 | 0.34 | 2.18 | 7.69 |
| MSE (mm) | / | / | 19.80 | 19.63 | 14.36 | 9.91 | 3.97 |
| $S$ (%) | / | / | 14.78 | 17.04 | 5.94 | 4.17 | 0.95 |
| $I$ (mm/s) | / | / | 33.52 | 53.56 | 116.20 | 49.8 | 0.90 |
| $b$ (rad) | -p | p | -0.27 | -1.60 | -1.28 | 0.31 | 0.52 |
| $g$ (rad) | -p | p | 1.88 | 2.95 | 3.07 | 3.14 | 3.26 |
| $x_a$ (mm) | -500 | 500 | 85.96 | -78.20 | -62.56 | 137.85 | 150.57 |
| $y_a$ (mm) | -500 | 500 | 126.16 | -66.32 | -79.58 | 84.31 | 66.12 |
| $r_1$ /mm | 30 | 500 | 40.94 | 75.31 | 90.38 | 49.63 | 56.20 |
| $r_2$ /mm | 30 | 500 | 341.58 | 431.32 | 498.06 | 78.12 | 88.15 |
| $r_3$ /mm | 30 | 500 | 253.82 | 379.86 | 303.89 | 87.94 | 99.62 |
| $r_4$ /mm | 30 | 500 | 147.64 | 382.04 | 458.45 | 102.20 | 128.33 |
| $r_5$ /mm | 30 | 500 | 64.54 | 136.34 | 136.62 | 50.25 | 51.20 |
| $r_6$ /mm | -500 | 500 | 49.21 | 40.55 | 32.44 | -6.13 | -4.21 |
| $r_7$ /mm | 30 | 500 | 159.03 | 328.87 | 263.10 | 86.79 | 96.50 |
| $r_8$ /mm | -500 | 500 | 283.92 | -398.41 | -328.33 | -21.20 | -25.13 |
| $r_9$ /mm | 30 | 500 | 355.97 | 391.06 | 459.06 | 142.52 | 127.65 |
| $r_{10}$ /mm | 30 | 500 | 199.21 | 387.16 | 309.73 | 101.39 | 83.21 |
| $r_{11}$ /mm | -500 | 500 | 0.00 | -5.75 | -4.60 | 4.96 | 8.09 |
| $r_{12}$ /mm | 30 | 500 | 179.44 | 243.03 | 234.60 | 129.56 | 115.54 |

Appendix C

Appendix C-1 Parameters of the optimization and selected individual in Section 4

|  | time /min | $h_6$ /mm | $h_4$ /mm | $r_1$ /mm | $r_2$ /mm | $r_3$ /mm | $r_4$ /mm | $r_5$ /mm | $r_6$ /mm | $x_d$ /mm | $y_d$ /mm | $a_1$ /mm | $b_1$ /mm | $a_3$ /mm | $b_3$ /mm | $\Delta x_{EH}$ /mm |
|---|---|---|---|---|---|---|---|---|---|---|---|---|---|---|---|---|
| $X_{lb}$ | / | / | / | 30 | 50 | 50 | 50 | 50 | 20 | -150 | -150 | 50 | 50 | 50 | 50 | 100 |
| $X_{ub}$ | / | / | / | 100 | 250 | 250 | 250 | 250 | 100 | 150 | 150 | 350 | 350 | 300 | 300 | 300 |
| case 1 | 13.3 | 50.1 | 220.1 | 40.2 | 106.1 | 75.8 | 99.0 | 36.5 | 26.5 | 90.9 | 98.7 | 155.3 | 128.8 | 87.1 | 102.3 | 113.6 |
| case 2 | 19.2 | 50.2 | 239.6 | 48.9 | 86.5 | 86.2 | 77.5 | 53.8 | 23.9 | 83.3 | 69.6 | 145.3 | 141.8 | 72.2 | 79.2 | 102.0 |
| case 3 | 34.3 | 50.0 | 260.1 | 43.5 | 140.5 | 155.2 | 110.9 | 23.5 | 47.8 | 69.0 | 54.7 | 223.1 | 247.2 | 139.9 | 85.5 | 182.7 |
| case 4 | 38.2 | 49.9 | 280.1 | 72.1 | 211.7 | 156.1 | 205.6 | 119.2 | 79.5 | 135.1 | 94.6 | 311.1 | 289.7 | 174.2 | 151.2 | 185.2 |
| case 5 | 31.9 | 49.9 | 300.5 | 73.6 | 197.0 | 152.1 | 183.3 | 124.7 | 72.3 | 142.2 | 99.8 | 305.5 | 285.6 | 180.8 | 163.4 | 167.1 |

Appendix C-2 Optimization results of different algorithms on Case 1 ($h_{6,t}$ = 50 mm, $h_{4,t}$ = 220 mm).

| Results | Case 1 | | | | | | |
|---|---|---|---|---|---|---|---|
|  | NSGA-II | CMOEMT | CMOES | DRLOS-EMCMO | IMTCMO | MCCMO | MOEA/D-CMT |
| $h_6$ (mm) | 50.05 | 50.00 | 49.94 | 49.97 | 50.05 | 49.76 | 49.99 |
| $h_4$ (mm) | 220.06 | 220.00 | 217.00 | 219.96 | 219.52 | 220.42 | 219.99 |
| $r_1$ (mm) | 40.15 | 46.68 | 55.33 | 38.54 | 60.54 | 70.75 | 38.01 |
| $r_2$ (mm) | 106.06 | 88.35 | 137.47 | 189.61 | 136.76 | 164.99 | 152.35 |
| $r_3$ (mm) | 75.76 | 70.31 | 140.15 | 79.34 | 113.66 | 117.07 | 145.28 |
| $r_4$ (mm) | 99.02 | 95.88 | 104.08 | 167.29 | 133.29 | 140.67 | 69.85 |
| $r_5$ (mm) | 36.52 | 17.19 | 157.99 | 4.96 | 78.20 | 141.73 | -31.06 |
| $r_6$ (mm) | 26.52 | 40.83 | 62.09 | 100.78 | 72.05 | 99.20 | 48.66 |
| $x_d$ (mm) | 90.91 | 93.74 | 105.71 | 126.94 | 54.78 | 50.71 | 97.42 |
| $y_d$ (mm) | 98.72 | 161.31 | 120.02 | 191.35 | 158.09 | 121.48 | 112.87 |
| $a_1$ (mm) | 155.31 | 175.60 | 287.30 | 231.01 | 227.85 | 271.74 | 247.94 |
| $b_1$ (mm) | 128.79 | 213.19 | 284.71 | 285.07 | 265.92 | 223.35 | 258.26 |
| $a_3$ (mm) | 87.12 | 117.12 | 140.10 | 161.21 | 118.92 | 157.40 | 96.38 |
| $b_3$ (mm) | 102.28 | 124.06 | 93.31 | 123.51 | 106.87 | 130.30 | 109.42 |
| $\Delta x_{EH}$ (mm) | 113.64 | 146.84 | 151.06 | 127.59 | 182.82 | 127.54 | 135.59 |

Appendix C-3 Optimization results of different algorithms on Case 2 ($h_{6,t}$ = 50 mm, $h_{4,t}$ = 240 mm).

| Results | Case 2 | | | | | | |
|---|---|---|---|---|---|---|---|
|  | NSGA-II | CMOEMT | CMOES | DRLOS-EMCMO | IMTCMO | MCCMO | MOEA/D-CMT |
| $h_6$ (mm) | 50.23 | 50.00 | 51.63 | 49.97 | 50.00 | 50.07 | 50.00 |
| $h_4$ (mm) | 239.64 | 240.00 | 238.37 | 240.48 | 239.97 | 240.88 | 239.99 |
| $r_1$ (mm) | 48.92 | 47.81 | 49.68 | 63.31 | 40.99 | 42.14 | 71.38 |
| $r_2$ (mm) | 86.51 | 130.14 | 170.08 | 152.36 | 99.55 | 141.90 | 125.48 |
| $r_3$ (mm) | 86.17 | 88.30 | 109.78 | 136.91 | 76.85 | 100.71 | 166.32 |
| $r_4$ (mm) | 77.49 | 124.95 | 180.02 | 135.30 | 124.92 | 191.71 | 90.51 |
| $r_5$ (mm) | 53.79 | 57.74 | 83.89 | 115.23 | 0.78 | 38.03 | 155.53 |
| $r_6$ (mm) | 23.89 | 47.30 | 91.62 | 60.79 | 78.33 | 95.19 | 50.67 |
| $x_d$ (mm) | 93.30 | 103.61 | 145.85 | 109.57 | 106.14 | 151.07 | 115.38 |
| $y_d$ (mm) | 69.62 | 134.97 | 190.70 | 139.62 | 122.31 | 193.27 | 170.08 |
| $a_1$ (mm) | 145.28 | 203.10 | 201.11 | 266.59 | 258.00 | 241.02 | 165.95 |
| $b_1$ (mm) | 141.78 | 214.64 | 267.39 | 255.90 | 280.00 | 279.58 | 275.92 |
| $a_3$ (mm) | 72.18 | 127.32 | 116.75 | 163.62 | 105.86 | 138.93 | 117.20 |
| $b_3$ (mm) | 79.17 | 115.10 | 126.12 | 91.51 | 103.60 | 124.77 | 122.09 |
| $\Delta x_{EH}$ (mm) | 101.95 | 144.18 | 192.54 | 99.28 | 188.05 | 147.42 | 199.45 |

Appendix C-4 Optimization results of different algorithms on Case 3 ($h_{6,t}$ = 50 mm, $h_{4,t}$ = 260 mm).

| Results | Case 3 | | | | | | |
|---|---|---|---|---|---|---|---|
|  | NSGA-II | CMOEMT | CMOES | DRLOS-EMCMO | IMTCMO | MCCMO | MOEA/D-CMT |
| $h_6$ (mm) | 50.00 | 50.07 | 50.27 | 50.00 | 49.91 | 49.95 | 49.97 |
| $h_4$ (mm) | 260.11 | 259.97 | 260.33 | 259.99 | 260.20 | 260.58 | 259.99 |
| $r_1$ (mm) | 43.51 | 61.50 | 65.90 | 44.06 | 39.03 | 73.30 | 52.76 |
| $r_2$ (mm) | 140.48 | 164.63 | 193.52 | 175.51 | 109.44 | 165.73 | 131.48 |
| $r_3$ (mm) | 155.24 | 133.13 | 103.57 | 178.05 | 77.84 | 151.36 | 94.33 |
| $r_4$ (mm) | 110.91 | 167.23 | 185.19 | 135.89 | 116.76 | 176.09 | 138.63 |
| $r_5$ (mm) | 23.59 | 93.06 | 80.21 | 24.20 | 49.21 | 122.35 | 54.02 |
| $r_6$ (mm) | 47.84 | 78.64 | 82.87 | 69.52 | 81.85 | 92.82 | 82.07 |
| $x_d$ (mm) | 69.00 | 125.64 | 64.33 | 177.29 | 98.04 | 169.65 | 121.22 |
| $y_d$ (mm) | 54.70 | 81.20 | 175.41 | 170.70 | 134.28 | 164.19 | 133.32 |
| $a_1$ (mm) | 223.12 | 239.41 | 189.70 | 269.56 | 289.53 | 210.73 | 158.69 |
| $b_1$ (mm) | 247.22 | 239.02 | 205.01 | 260.37 | 255.16 | 202.51 | 169.84 |
| $a_3$ (mm) | 139.91 | 89.39 | 150.56 | 165.10 | 287.02 | 163.09 | 139.79 |
| $b_3$ (mm) | 85.47 | 181.46 | 113.19 | 179.17 | 125.28 | 147.41 | 122.75 |
| $\Delta x_{EH}$ (mm) | 182.71 | 146.80 | 151.06 | 161.69 | 142.06 | 113.84 | 170.73 |

Appendix C-5 Optimization results of different algorithms on Case 4 ($h_{6,t}$ = 50 mm, $h_{4,t}$ = 280 mm).

| Results | Case 4 | | | | | | |
|---|---|---|---|---|---|---|---|
| | NSGA-II | CMOEMT | CMOES | DRLOS-EMCMO | IMTCMO | MCCMO | MOEA/D-CMT |
| $h_6$ (mm) | 49.91 | 50.04 | 49.80 | 50.00 | 50.03 | 49.94 | 49.97 |
| $h_4$ (mm) | 280.10 | 280.03 | 281.52 | 280.00 | 279.93 | 280.13 | 280.19 |
| $r_1$ (mm) | 72.07 | 57.29 | 60.05 | 74.35 | 45.43 | 47.16 | 31.54 |
| $r_2$ (mm) | 211.70 | 159.17 | 136.74 | 173.32 | 192.31 | 147.69 | 64.99 |
| $r_3$ (mm) | 156.08 | 177.16 | 109.01 | 159.24 | 89.74 | 140.18 | 116.90 |
| $r_4$ (mm) | 205.60 | 144.79 | 134.90 | 159.92 | 195.02 | 123.65 | 117.70 |
| $r_5$ (mm) | 119.22 | 47.98 | 104.61 | 129.82 | 53.16 | -33.19 | 78.07 |
| $r_6$ (mm) | 79.53 | 51.69 | 67.55 | 92.58 | 94.06 | 28.04 | 83.13 |
| $x_d$ (mm) | 135.12 | 80.25 | 95.85 | 157.40 | 123.86 | 89.53 | 128.06 |
| $y_d$ (mm) | 94.58 | 89.56 | 153.52 | 143.39 | 186.88 | 155.12 | 88.57 |
| $a_1$ (mm) | 311.12 | 251.54 | 220.92 | 254.28 | 280.52 | 205.63 | 217.47 |
| $b_1$ (mm) | 289.75 | 259.93 | 224.49 | 232.42 | 327.35 | 241.91 | 204.24 |
| $a_3$ (mm) | 174.21 | 158.52 | 140.57 | 151.36 | 202.27 | 117.74 | 93.96 |
| $b_3$ (mm) | 151.20 | 145.25 | 133.17 | 129.42 | 120.47 | 136.22 | 124.44 |
| $\Delta x_{EH}$ (mm) | 185.19 | 175.52 | 141.72 | 140.31 | 130.12 | 175.43 | 206.25 |

Appendix C-6 Optimization results of different algorithms on Case 5 ($h_{6,t}$ = 50 mm, $h_{4,t}$ = 300 mm).

| Results | Case 5 | | | | | | |
|---|---|---|---|---|---|---|---|
| | NSGA-II | CMOEMT | CMOES | DRLOS-EMCMO | IMTCMO | MCCMO | MOEA/D-CMT |
| $h_6$ (mm) | 49.88 | 49.89 | 49.06 | 299.78 | 50.03 | 49.23 | 49.99 |
| $h_4$ (mm) | 300.49 | 300.01 | 301.29 | 50.13 | 300.74 | 300.85 | 300.00 |
| $r_1$ (mm) | 73.57 | 59.33 | 32.95 | 58.37 | 37.86 | 48.02 | 47.88 |
| $r_2$ (mm) | 197.03 | 140.78 | 115.48 | 115.79 | 145.98 | 205.29 | 129.40 |
| $r_3$ (mm) | 152.13 | 91.23 | 73.60 | 120.54 | 109.39 | 106.99 | 105.14 |
| $r_4$ (mm) | 183.31 | 146.54 | 148.33 | 116.03 | 113.42 | 198.56 | 123.93 |
| $r_5$ (mm) | 124.70 | 64.23 | 27.12 | 87.94 | -33.72 | 100.25 | 83.31 |
| $r_6$ (mm) | 72.33 | 58.59 | 27.83 | 97.92 | 90.44 | 25.84 | 89.81 |
| $x_d$ (mm) | 142.16 | 72.34 | 140.71 | 127.39 | 159.02 | 172.86 | 53.72 |
| $y_d$ (mm) | 99.78 | 112.56 | 176.58 | 153.07 | 61.97 | 142.95 | 88.34 |
| $a_1$ (mm) | 305.52 | 213.17 | 284.23 | 271.97 | 300.52 | 288.34 | 226.53 |
| $b_1$ (mm) | 285.56 | 216.62 | 229.38 | 240.75 | 276.34 | 288.33 | 156.13 |
| $a_3$ (mm) | 180.82 | 101.21 | 189.66 | 157.42 | 157.24 | 165.98 | 105.36 |
| $b_3$ (mm) | 163.36 | 168.28 | 155.37 | 143.95 | 151.07 | 150.52 | 156.50 |
| $\Delta x_{EH}$ (mm) | 167.10 | 164.33 | 160.01 | 120.83 | 180.16 | 184.39 | 206.48 |